\documentclass[preprint,journal,twoside]{IEEEtran}
\usepackage[table]{xcolor}
\usepackage{cite}
\usepackage{dsfont}
\usepackage{amsmath,amssymb,amsfonts}
\usepackage{bm}
\usepackage{float}
\usepackage{caption}
\usepackage{subcaption}
\usepackage{setspace}
\usepackage{algorithm}

\usepackage{amsthm}
\usepackage{algpseudocode}
\usepackage{bigints}
\usepackage{relsize}
\usepackage{braket}
\usepackage{graphicx}
\usepackage{textcomp}
\usepackage{stmaryrd}
\usepackage{array,booktabs,,dcolumn,caption}
\usepackage{tabularx}
\usepackage[colorlinks=true, allcolors=blue]{hyperref}

\def\BState{\State\hskip-\ALG@thistlm}
\theoremstyle{definition}
\newtheorem{definition}{Definition}[section]

\newcolumntype{d}[1]{D{.}{.}{#1}} 
\makeatletter
\newcolumntype{B}[3]{>{\boldmath\DC@{#1}{#2}{#3}}c<{\DC@end}}
\renewcommand{\ast}{{}^{\textstyle *}} 
\renewcommand{\vec}[1]{\mathbf{#1}}

\newcommand{\fn}{f(\mathbf{r})}
\newcommand{\fa}{f_a(\mathbf{r})}

\newcommand{\faHMmeas}{\hat{f}_{a,meas}^{HM}(\mathbf{r})}
\newcommand{\faHMnull}{\hat{f}_{a,null}^{HM}(\mathbf{r})}
\newcommand{\fameas}{f_{a,meas}(\mathbf{r})}
\newcommand{\fanull}{f_{a,null}(\mathbf{r})}

\newcommand{\hfa}{\hat{f}_a(\mathbf{r})}

\newcommand{\expn}{\psi_n(\vec{r})}
\newcommand{\coeff}{\bm{\theta}}
\newcommand{\cmeas}{\coeff_{meas}}
\newcommand{\hcmeas}{\hat{\coeff}_{meas}}
\newcommand{\cnull}{\coeff_{null}}
\newcommand{\hcnull}{\hat{\coeff}_{null}}
\newcommand{\hcoeff}{\hat{\bm{\theta}}}
\newcommand{\hcpinv}{\hat{\coeff}_{tp}}

\newcommand{\cHMnull}{\hat{\coeff}^{HM}_{null}}

\newcommand{\MP}{\vec{H}^+}

\newcommand{\Umeas}{\mathcal{N}^{\perp}_{P}(\vec{H})}
\newcommand{\Unull}{\mathcal{N}_{P}(\vec{H})}
\newcommand{\spaceU}{\mathbb{L}_2(\mathbb{R}^d)}
\newcommand{\sv}{\sqrt{\mu_n}}

\newcommand{\clearsubcaptcounter}{\setcounter{sub\@captype}{0}}

\DeclareMathOperator*{\argmin}{argmin}
\DeclareMathOperator*{\argmax}{argmax}

\DeclareMathAlphabet\mathbfcal{OMS}{cmsy}{b}{n}
\def\BibTeX{{\rm B\kern-.05em{\sc i\kern-.025em b}\kern-.08em
    T\kern-.1667em\lower.7ex\hbox{E}\kern-.125emX}}
\begin{document}

\title{On hallucinations in tomographic \\ image reconstruction}

\author{Sayantan Bhadra*, Varun A. Kelkar*, Frank J. Brooks and Mark A. Anastasio,~\IEEEmembership{Senior Member,~IEEE}
\thanks{This work was supported in part by NIH Awards EB020604, EB023045, NS102213, EB028652, and NSF Award DMS1614305.}
\thanks{Sayantan Bhadra is 
with the Department of Computer Science and Engineering, Washington University in St. Louis, St. Louis, 
MO 63130 USA (e-mail: sayantanbhadra@wustl.edu).}
\thanks{Varun A. Kelkar is with 
the Department of Electrical and Computer Engineering, University of Illinois at Urbana-Champaign, Urbana, IL 61801 USA (e-mail: vak2@illinois.edu).}
\thanks{Frank J. Brooks and Mark A. Anastasio are with 
the Department of Bioengineering, University of Illinois at Urbana-Champaign, Urbana, IL 61801 USA (e-mail: {fjb,maa}@illinois.edu).}
\thanks{*Sayantan Bhadra and Varun A. Kelkar contributed equally.}
}
\maketitle

\begin{abstract}
Tomographic image reconstruction is generally an ill-posed linear inverse problem. Such ill-posed inverse problems are typically regularized using prior knowledge of the sought-after object property. Recently, deep neural networks have been actively investigated for regularizing image reconstruction problems by learning a prior for the object properties from training images. However, an analysis of the prior information learned by these deep networks and their ability to generalize to data that may lie outside the training distribution is still being explored. An inaccurate prior might lead to false structures being hallucinated in the reconstructed image and that is a cause for serious concern in medical imaging. In this work, we propose to illustrate the effect of the prior imposed by a reconstruction method by
decomposing the image estimate into generalized measurement and null components. The concept of a hallucination map is introduced for the general purpose of understanding the effect of the prior in regularized reconstruction methods. Numerical studies are conducted corresponding to a stylized tomographic imaging modality. The behavior of different reconstruction methods under the proposed formalism is discussed with the help of the numerical studies.

\end{abstract}


\section{Introduction}
\label{sec:introduction}
\IEEEPARstart{I}n tomographic imaging, a reconstruction method is employed to estimate the sought-after object from a collection of measurements obtained from an imaging system \cite{kak2002principles}. Since the sought-after object is usually described as a continuous function and the measurements are discrete, image reconstruction methods usually seek a finite-dimensional estimate of the object. Moreover, it is often desirable to reconstruct images from as few measurements as possible, without compromising on the diagnostic quality of the image. For example, data-acquisition times in magnetic resonance imaging (MRI) can be reduced by undersampling the k-space  \cite{yang2016sparse}. In such situations the acquired measurements are said to be \textit{sparse}, i.e., they are generally insufficient to uniquely specify a finite-dimensional approximation of the sought-after object,
even in the absence of measurement noise or errors related to modeling the imaging system. This naturally implies that the inverse problem is ill-posed and some form of regularization needs to be performed with priors imposed on the sought-after object. Various methods have been proposed for regularization that can effectively mitigate the impact of measurement-incompleteness on image reconstruction. Among these methods, regularization using sparsity-promoting penalties has been employed widely \cite{donoho2006most, donoho2003optimally, candes2006stable, sidky2008image}.

Recently, there has been considerable focus on developing regularization strategies that seek to learn the prior distribution that describes the object to-be-imaged from existing data, instead of using hand-crafted priors such as sparsity-promoting penalties. Nascent deep learning-based methods have inspired a new wave of reconstruction methods that implicitly or explicitly learn the prior distribution from a set of training images in order to regularize the reconstruction problem \cite{wang2016perspective,mccann2017convolutional,ravishankar2019image}. However, such learning-based methods have also raised concerns regarding their robustness \cite{huang2018some,gottschling2020troublesome,antun2020instabilities,laves2020uncertainty} and their ability to generalize to measurements that may lie outside the distribution of the training data \cite{antun2020instabilities,asim2020invertible,kelkar2021compressible}. This is particularly relevant in the field of medical imaging where novel abnormalities can be present in the observed measurement data that may not be encountered even with a large training dataset. Moreover, simulation studies have shown that deep learning-based reconstruction methods are inherently unstable, i.e. small perturbations in the measurement may produce large differences in the reconstructed image \cite{gottschling2020troublesome,antun2020instabilities}. 

The potential lack of generalization of deep learning-based reconstruction methods as well as their innate unstable nature may cause false structures to appear in the reconstructed image that are absent in the object being imaged. These false structures may arise due to the reconstruction method incorrectly estimating parts of the object that either did not contribute to the observed measurement data or cannot be recovered in a stable manner, a phenomenon that can be termed as \textit{hallucination}. The presence of such false structures in reconstructed images can possibly lead to an incorrect medical diagnosis. Hence, there is an urgent need to investigate the nature and impact of false structures arising out of hallucinations from deep learning-based reconstruction methods for tomographic imaging.

The topic of image hallucinations has previously been studied within the context of image super-resolution \cite{baker2002limits,liu2005hallucinating,wang2014comprehensive,fawzi2016image}. 
In image super-resolution, the term hallucination generally refers to high-frequency features that are introduced into the high-resolution image but do not exist in the measured low-resolution image. Hallucinations can also be realized in more general inverse problems such as image reconstruction. In such cases, the structure of the imaging operator null space is generally more complicated and the hallucinations may not be confined to high-frequency structures \cite{barrett2013foundations}. However, a formal definition of hallucinations within the context of such inverse problems has not been reported.

This study proposes a way to mathematically formalize the concept of hallucinations for general linear imaging systems that is consistent with both the mathematical notion of a hallucination in image super-resolution and the intuitive notion of hallucinations as ``artifacts or incorrect features that occur due to the prior that cannot be produced from the measurements". In addition, the notion of a \textit{task-informed} or \textit{specific} hallucination map is introduced. Through preliminary numerical studies, the behavior of different reconstruction methods under the proposed formalism is illustrated. It is shown that, in certain cases, traditional error maps are insufficient for visualizing and detecting specific hallucinations.


The remainder of this paper is organized as follows. In Sec. \ref{sec:Background}, salient aspects of linear operator theory are reviewed, and the need for describing hallucinations based on the measurement and null space components is motivated. The concept of a hallucination map is introduced in Sec. \ref{sec:definition}, along with a definition of specific hallucination maps. Sections \ref{sec:Numerical studies} and \ref{sec:results} describe the numerical studies performed to demonstrate the potential utility of proposed hallucination maps with a stylized tomographic imaging modality. Finally, a discussion and summary of the work is presented in Sec. \ref{sec:discussion}.

\section{Background}
\label{sec:Background}
\subsection{Imaging models}
A linear digital imaging system can be described as a continuous-to-discrete (CD) mapping \cite{barrett2013foundations,anastasio2003basic}:
\begin{equation}\label{eq:imaging}
\mathbf{g} = \mathbfcal{H}\fn + \vec{n},
\end{equation}
where $\fn \in \spaceU$ is a function of continuous variables that represents the object being imaged, the vector $\vec{g} \in \mathbb{E}^M$ denotes the measured data samples and $\vec{n} \in \mathbb{E}^M$ is the measurement noise. The linear CD operator $\mathbfcal{H}:\spaceU \rightarrow \mathbb{E}^M$  describes the action of the imaging system. In practice, discrete-to-discrete (DD) imaging models are often employed as approximations to the true CD imaging model. In a DD model, an $N$-dimensional approximation of $\fn$ is utilized \cite{barrett2013foundations,anastasio2003basic}:
\begin{equation}\label{eq:DD}
\fa = \sum_{n=1}^N [\theta]_n \expn,
\end{equation}
where the subscript $a$ stands for approximate, $[\theta]_n$ is the $n$-th element of the coefficient vector $\bm{\theta} \in \mathbb{E}^N$ and $\expn$ is the $n$-th expansion function.  On substitution from Eq. \eqref{eq:DD} in Eq. \eqref{eq:imaging}, the DD imaging system can be expressed as 
\begin{equation}\label{eq:imaging_DD}
\vec{g} \approx \mathbfcal{H} \fa + \vec{n}= \sum_{n=1}^N [\theta]_n \mathbfcal{H} \expn + \vec{n}\equiv \vec{H}\coeff + \vec{n},
\end{equation}
where $\vec{H}:\mathbb{E}^N \rightarrow \mathbb{E}^M$ is the system matrix constructed using $\mathbfcal{H}$ and $\{\expn\}_{n=1}^N$. Image reconstruction methods based on Eq.\ (\ref{eq:imaging_DD}) seek to estimate $\coeff$ from $\vec{g}$, after which the approximate object function $\fa$ can be determined by use of Eq. \eqref{eq:DD}. A well-known expansion function is the pixel expansion function. For two-dimensional objects $\fn$ with $\vec{r} = (x,y)$, the pixel expansion function can be expressed as  \cite{anastasio2003basic}:
\begin{equation}
\psi_n(\vec{r}) = 
\begin{cases}
1,& \text{if  } |x-x_n| \text{ and } |y-y_n| \leq \frac{\gamma}{2}\\
0,& \text{otherwise}
\end{cases}
\end{equation}
where  $\vec{r}_n=(x_n,y_n)$ represents the coordinates of the $n$-th grid point of a uniform Cartesian lattice and $\gamma$ denotes the spacing between the lattice points. When a pixel expansion function is employed, the corresponding coefficient vector $\coeff$ directly provides a digital image representation of the continuous object function $\fa$.

\subsection{Generalized measurement and null components}
\label{sec:meas_null_functions}
For the DD imaging model described by Eq. \eqref{eq:imaging_DD}, the properties of $\vec{H}$ affect the ability to estimate $\coeff$ uniquely and stably. In the absence of measurement noise, $\coeff$ can be determined uniquely from measurements $\vec{H}\coeff$ when $\vec{H}$ is injective or if $\coeff$ is known to lie in a subset $S$ of $\mathbb{E}^N$ and the restriction $\vec{H}\left|_S\right.$ is injective.
The ability to stably reconstruct an estimate of $\coeff$ 
is also of fundamental importance.
\color{black}
Stability is a way of quantifying how ``close'' two estimates $\hat{\coeff}_1, \hat{\coeff}_2$ of $\coeff$ will be, if they are estimated from two ``close'' measurement vectors $\mathbf{g}_1$ and $\mathbf{g}_2$ respectively. For instance, $\vec{g}_1$ and $\vec{g}_2$ may correspond to the same object but differ due to them having two different measurement noise realizations. A popular notion of stability is based on how the $\ell_2$-distance between $\hat{\coeff}_1$ and $\hat{\coeff}_2$ relates to that between $\vec{g}_1$ and $\vec{g_2}$ \cite{bal2012introduction}:
\begin{equation}\label{eq:stability}
\| \hat{\coeff}_1 - \hat{\coeff}_2\|_2 \leq \alpha\| \vec{g}_1 - \vec{g}_2 \|_2,
\end{equation}
where $\alpha$ is a constant that is additionally required to be smaller than a tolerance value $\epsilon$.


The ability to estimate $\coeff$ stably can be analyzed through the lens of the singular value decomposition (SVD) of $\vec{H}$ \cite{barrett2013foundations}:
\begin{equation}\label{eq:SVD}
\vec{H} = \sum_{n=1}^{R} \sqrt{\mu_n}\vec{v}_n \vec{u}_n^{\dagger}.
\end{equation}
Here, $\vec{u}_n$ and $\vec{v}_n$ are the singular vectors of $\vec{H}$ and $\sv$ are the singular values. The vector $\vec{u}_n^{\dagger}$ is the adjoint of $\vec{u}_n$ and the integer $R > 0$ denotes the rank of $\vec{H}$, where $\vec{H}$ is not necessarily full-rank. The singular values $\sv$ are ordered such that $\mu_1\geq\mu_2\geq \cdots \geq \mu_R > 0$. 

A  pseudoinverse-based estimate of $\coeff$ can be computed as $\hat{\coeff}_{pinv} \equiv \vec{H}^+ \vec{g}$, where the linear operator $\vec{H}^+$ denotes the Moore-Penrose pseudoinverse of $\vec{H}$ that can be expressed as
\begin{equation}\label{eq:MP_eq}
\MP = \sum_{n=1}^R \frac{1}{\sv} \vec{u}_n \vec{v}_n^{\dagger}.
\end{equation}

From Eq. \eqref{eq:imaging_DD}, due to the linearity of $\MP$, $\hat{\coeff}_{pinv}$ can be represented as
\begin{equation}\label{eq:pinv_noise}
    \hat{\coeff}_{pinv} = \MP \vec{g} \approx \MP (\vec{H}\coeff+\vec{n}) = \MP\vec{H}\coeff + \MP \vec{n}.
\end{equation}

Due to the presence of the term $\MP \vec{n}$ in Eq. \eqref{eq:pinv_noise}, when the trailing singular values of $\vec{H}$ are small, $\alpha$ in Eq. \eqref{eq:stability} is large, leading to unstable estimates of $\coeff$. In this scenario, a truncated pseudoinverse can be defined as
\begin{equation}\label{eq:MP_eq_eps}
\MP_{P} = \sum_{n=1}^P \frac{1}{\sv} \vec{u}_n \vec{v}_n^{\dagger},
\end{equation}
where the integer $P\le R$ is chosen such that, for a given tolerance $\epsilon$, $\MP_P \vec{g}$ is a stable, linear estimate of $\coeff$ according to Eq. \eqref{eq:stability} with $\mu_P > 1/\epsilon^2 \geq \mu_{P+1}$.
The truncated pseudoinverse can be used to form projection operators that project $\coeff \in \mathbb{E}^N$ onto orthogonal subspaces -- the `generalized' null  and measurement spaces \cite{deal1996nullspace}. The generalized null space of $\vec{H}$, denoted by $\Unull$, is spanned by the singular vectors $\{\vec{u}_n\}_{n=P+1}^N$ that correspond to singular values satisfying $\sv \leq 1/\epsilon$. The orthogonal complement of the generalized null space  is the generalized measurement space $\Umeas$.

\begin{definition}[\textit{Generalized measurement and null components}]

    Let $\vec{H}$ and $\vec{H}^+_P$ denote the forward and truncated pseudoinverse operators,
    described in Equations \eqref{eq:imaging_DD} and \eqref{eq:MP_eq_eps} respectively.
    Let $\vec{H}_P$ denote the truncated forward operator, defined as
    \begin{align}
    \vec{H}_P = \sum_{n=1}^{P} \sqrt{\mu_n}\vec{v}_n \vec{u}_n^{\dagger}.
    \end{align}
    Note that $\vec{H}_P^+ = (\vec{H}_P)^+$.
    \color{black}
    The coefficient vector $\coeff$ can be uniquely decomposed as $\coeff = \coeff_{meas} + \coeff_{null}$, where $\coeff_{meas} \in \Umeas$ and $\coeff_{null} \in \Unull$ are specified as
    \begin{equation}\label{eq:project_meas_MP}
    \cmeas = \mathcal{P}_{meas}\coeff = \MP_P \vec{H}\coeff = \MP_P \vec{H}_P\coeff,
    \end{equation}
    and
    \begin{equation}\label{eq:project_null_MP}
    \cnull = \mathcal{P}_{null}\coeff = [\vec{I}_N - \MP_P \vec{H}]\coeff = [\vec{I}_N - \MP_P \vec{H}_P]\coeff.
    \end{equation}    
    \color{black}
    Here, the projection operators $\mathcal{P}_{meas}$ and $\mathcal{P}_{null}$ project $\coeff$ to $\Umeas$ and $\Unull$ \cite{barrett2013foundations}, and $\vec{I}_N$ is the identity operator in $\mathbb{E}^N$. 
\end{definition}

In special cases where the singular values $\sv$ and the tolerance $\epsilon$ are such that $P=R$, the generalized null space is spanned by the singular vectors $\{\vec{u}_n\}_{n=R+1}^N$ with singular values $\sv = 0$. In such cases, the generalized null space reduces to the true null space
\begin{equation}\label{eq:null_space}
\Unull = \mathcal{N}(\vec{H}) \equiv \{\coeff \in \mathbb{E}^N | \vec{H}\coeff = \mathbf{0}\},
\end{equation}
where $\mathbf{0}$ is the zero vector in $\mathbb{E}^M$. Correspondingly, the true measurement space is the orthogonal complement of the true null space.  By definition, the true null space contains those object vectors that are mapped to the zero measurement data vector and hence are `invisible'  to the imaging system.

Having obtained the generalized measurement and null components of $\coeff$, the approximate object function $\fa$ can also be decomposed into generalized measurement and null components:
\begin{align} \label{eq:meas_null_fn}
\fa &= \sum_{n=1}^N [\theta]_n \expn \nonumber \\
&= \sum_{n=1}^N [\theta_{meas}]_n \expn + \sum_{n=1}^N [\theta_{null}]_n \expn \nonumber\\
&= \fameas + \fanull.
\end{align}

Note that for all $\vec{g}_1, \vec{g}_2 \in \mathbb{E}^M$, $\| \MP_P \vec{g}_1 - \MP_P \vec{g}_2 \| \leq (1/\sqrt{\mu_P}) \| \vec{g}_1  - \vec{g_2}\|$, whereas for all $\sigma \in \Unull$, 
$\|\sigma \| \geq \|\vec{H}\sigma\|/\sqrt{\mu_{P+1}}$.
Hence, for a given $\coeff \in \mathbb{E}^N$, $\coeff_{meas}$ is the component of $\coeff$ that can be stably estimated via the truncated pseudoinverse from the measurement data. Contrarily, $\coeff_{null}$ cannot be stably estimated from the measurement data alone; additional information provided through priors and regularization is needed to estimate this component. These observations will be essential to the definitions of hallucinations that are provided later.

\vspace{-0.3cm}
\subsection{Regularization in tomographic image reconstruction}
\label{sec:recon}
As discussed above, in order to obtain a stable estimate of $\coeff$ from incomplete and/or noisy measurements, imposition of prior knowledge about the object is generally needed. A flexible method of incorporating priors in the estimation of $\coeff$ is through the Bayesian formalism, where $\coeff$, $\vec{g}$ and $\vec{n}$ are treated as instances of random variables with distributions $p_{\theta}$, $p_{\vec{\sf{g}}}$ and $p_{\vec{\sf{n}}}$ respectively \cite{bal2012introduction}. It is assumed that $p_{{\theta}}$, i.e. the distribution over all objects is known, and is 
called the \textit{prior}. By Bayes' theorem, the posterior distribution $p_{\coeff | \vec{\sf{g}}}$, given by 
\begin{equation}\label{eq:bayes}
p_{\coeff | \vec{\sf{g}}}(\coeff | \vec{g}) = \frac{p_{{\sf{g}} | \theta}(\vec{g} | \coeff) p_{\theta}(\coeff)} {p_{\sf{g}}(\vec{g})},
\end{equation}
characterizes the probability density over all possible values of the object given the prior and the noise model. Estimates such as the \textit{maximum a posteriori} (MAP) estimate $\argmax_{\coeff} p_{\coeff | \vec{\sf{g}}}(\coeff | \vec{g})$ can then be obtained from the posterior.

Regularization via penalization is an alternative formalism to incorporate prior knowledge. Here, the image reconstruction task is formulated as an optimization problem such as \cite{anastasio2003basic}
\begin{equation}\label{eq:optim}
\hat{\coeff} = \argmin_{\coeff} C_d(\vec{g},\vec{H}\coeff) + \lambda C_p(\coeff),
\end{equation}
where the data fidelity term $C_d(\vec{g},\vec{H}\coeff)$ enforces the estimate $\hat{\coeff}$ when acted upon by $\vec{H}$ to agree with the observed measurement data $\vec{g}$ and the penalty term $C_p (\coeff)$ encourages the solution to be consistent with the assumed prior. The hyper-parameter $\lambda$ controls the trade-off between data fidelity and regularization. Often, the penalty term $R (\coeff)$ is hand-crafted to encode priors such as the smoothness of natural images or sparsity of natural images in some transform domain \cite{ravishankar2019image}. The solution obtained through this formalism can be interpreted as the MAP estimate obtained from the Bayesian formalism in Eq. \eqref{eq:bayes}, with $p_{\coeff}(\coeff) = \text{exp}(-\lambda C_p( \coeff ))$ and $p_{{\sf{g}} | \coeff }(\vec{g} | \coeff) = \text{exp}(-C_d(\vec{g},\vec{H}\coeff))$.

Regularization can also be interpreted as restricting the possible solutions to a subset $S_{\mu} \subset \mathbb{E}^N$, with $S_{\mu}$ being a member of a family of subsets parameterized by $\mu$. The reconstruction procedure can then be represented by a possibly nonlinear mapping $\mathcal{R}_{\mu} : \mathbb{E}^M \rightarrow S_{\mu}$, with the image estimate given by $\hat{\coeff} = \mathcal{R}_{\mu}(\vec{g})$. Ideally, it is desirable that $\mathcal{R}_{\mu}$ satisfies the stability criterion described in Eq. \eqref{eq:stability}. 

\begin{figure*}[!htb]
    \centering
     \includegraphics[width=1\linewidth]{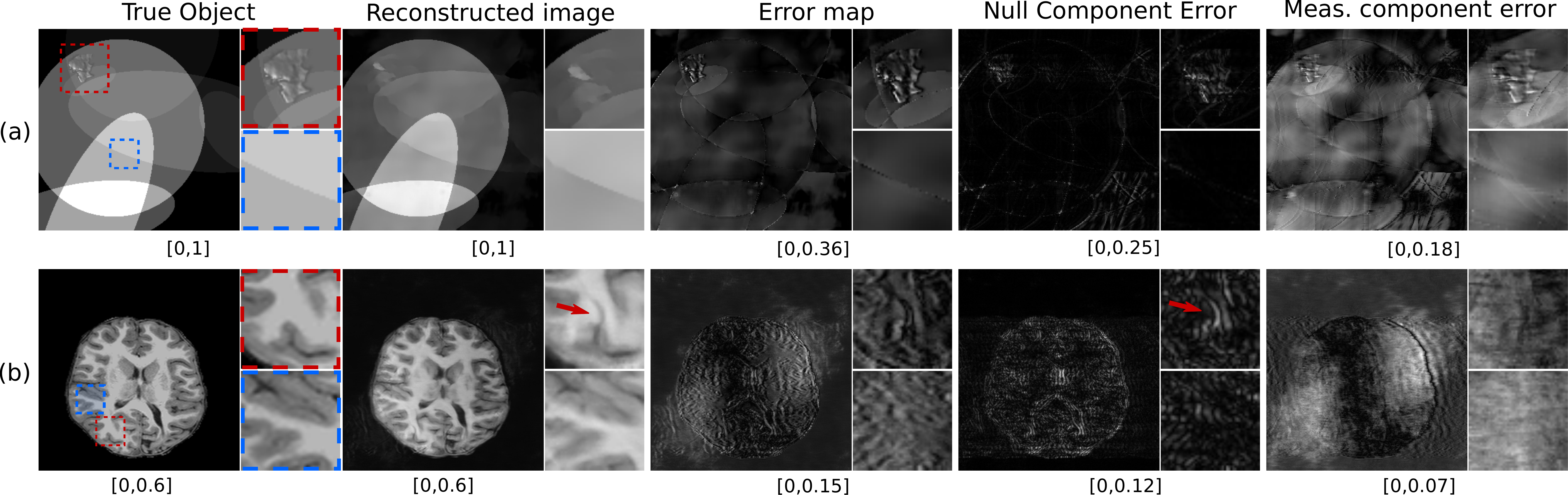}
      \caption{From left-to-right are examples of a true object, a reconstructed estimate of the object produced by use of a U-Net from tomographic measurements,  the total error map, the error in the null component of the reconstructed object, and the error in the measurement component of the reconstructed object. The two columns correspond to different objects.
      In each case, the true object is outside the respective training data distribution of the U-Net and phase noise was added to the measurements prior to image reconstruction.}
      \vspace{-0.5cm}
      \label{fig:motivating_image}
\end{figure*}

Recently, methods that implicitly learn a regularizer from existing data have been proposed. Methods based on dictionary learning and learning sparsifying transforms were some of the earliest applications of such data-driven regularization \cite{ravishankar2010mr,tiwari2019study,ravishankar2012learning,ravishankar2016data}. However, the most actively investigated data-driven regularization methods involve learning from training data by use of deep neural networks, popularly known as deep learning \cite{goodfellow2016deep,wang2016perspective}. 
Deep learning has been employed in different ways to explicitly or implicitly impose priors in  image reconstruction problems. For example, within the context of an end-to-end learned reconstruction mapping, a prior is imposed that is implicitly specified by the distribution of training data and network topology.
A comprehensive survey of the current state of deep learning-based methods in tomographic image reconstruction can be found in recent reviews, \cite{ravishankar2019image, mccann2019biomedical, hammernik2020machine}.

However, there have been growing concerns regarding the ability of data-driven and learning based reconstruction methods to generalize to measurements that lie outside the training distribution \cite{stayman2012information,huang2018some,gottschling2020troublesome,antun2020instabilities}. Moreover, deep learning-based reconstruction methods have been shown to not be uniformly stable, in the sense that certain imperceptible perturbations in the measurements may lead to large fluctuations in the reconstructed estimate \cite{gottschling2020troublesome,antun2020instabilities}. Such phenomena may lead to false structures appearing in the reconstructed image that do not exist in the object being imaged, and cannot be recovered stably from the original measurement data.
\vspace*{-0.2cm}
\section{Definition of hallucination maps}
\label{sec:definition}

When comparing or evaluating image reconstruction methods, it may be useful to visualize and quantify false structures that cannot be stably reconstructed from the measurements. Such structures have been colloquially referred to as being `hallucinated' and are attributable to use of an imperfect reconstruction prior.  
Error maps that display the difference between the reconstructed image estimate and the true object are commonly employed to assess reconstruction errors. 
Artifacts  revealed by error maps encompass a broad range of deviations that can appear in a reconstructed image with respect to its depiction of the object function being imaged. For example, incorrect modeling of the system matrix $\vec{H}$ or  measurement noise can lead to artifacts.
Consequently, as demonstrated in Fig. \ref{fig:motivating_image}, it may not be possible to isolate and label the artifacts attributable to the reconstruction prior from the error map alone.  A possible way to circumvent this is to compute separate error maps for the null and measurement components of the reconstructed image estimate. However, precise definitions for hallucinations in these sub-spaces have been  lacking.

In order to visualize and quantify hallucinations in tomographic images, measurement and null space hallucination maps are formally defined below. The proposed definitions are general and can be applied to analyze hallucinations produced by any reconstruction method that seeks to invert a linear imaging model.
The defined hallucination maps will permit isolation of image artifacts that cannot be stably reconstructed from the measurement data and
are attributable to the implicit or explicit  reconstruction prior.

\subsection{Hallucination map in the generalized measurement space}
\label{sec:meas_h_map}
Let $\hcoeff$ denote the estimate of the coefficient vector $\coeff$ obtained from $\vec{g}$ by use of an image reconstruction method. 
It is desirable that the projection of $\hat{\coeff}$ onto the generalized measurement space $\Umeas$, i.e. $\hcmeas$, should be near the truncated pseudoinverse solution $\hcpinv \equiv \MP_P \vec{g}$. This would ensure that $\hcmeas$ is consistent with the estimate of $\coeff$ that can be stably recovered from $\vec{g}$. However, due to the imposed regularization in a reconstruction method, there may be discrepancies in $\hcmeas$ with respect to the stable estimate $\hcpinv$ in the generalized measurement space $\Umeas$. In order to quantify such differences, a hallucination map in the generalized measurement space is defined as follows.

\begin{definition}[\textit{Generalized measurement space hallucination map}]
    As previously defined, let $\hat{\coeff}$ be an image estimate obtained by use of a reconstruction method and let $\hcpinv$ be the truncated pseudoinverse solution. The hallucination map in the measurement space is defined as,
    \begin{equation}\label{eq:hal_map_meas}
    \hcmeas^{HM} \equiv \hcmeas-\hcpinv.
    \end{equation}
            
\end{definition}
It should be noted that the computation of the hallucination map in the generalized measurement space requires no knowledge of the true object and simply reveals errors in the measurement component of $\hat{\coeff}$ with respect to the stably computed estimate $\hcpinv$.


For use in cases where pixel expansion functions are not employed, it is useful to translate the definition of hallucination maps to the subspace of the object space $\spaceU$ spanned by a generic basis $\{\psi_n(\mathbf{r})\}_{i=1}^N$.
\color{black}
By use of Eq. \eqref{eq:DD}, the estimate of $\fa$ can be represented as
\begin{equation}\label{eq:DD_est}
\hfa = \sum_{n=1}^N [\hat{\theta}]_n \expn.
\end{equation}
The hallucination map $\faHMmeas$ can be defined in the space $\spaceU$ as
\begin{equation}\label{eq:hal_map_meas_CD}
\faHMmeas \equiv \sum_{n=1}^N [\hat{\theta}^{HM}_{meas}]_n \expn.
\end{equation}

\subsection{Hallucination map in the generalized null space}
\label{sec:null_h_map}
As reviewed in Sec. \ref{sec:meas_null_functions}, to estimate the generalized null vector $\cnull$ from $\vec{g}$, reconstruction methods that impose appropriate priors are required. Hence, to accurately capture the effect of the prior on the reconstructed image, a definition of hallucinations must satisfy the following two desiderata:
\begin{itemize}
    \item The definition must involve the assessment of how accurate the estimate $\hat{\coeff}_{null} = \mathcal{P}_{null}\hat{\coeff}$ is as compared to the true generalized null vector $\cnull$.
    \item Since no prior is used in obtaining $\hcpinv$, the definition must ensure that $\hcpinv$ does not have any null space hallucinations.
\end{itemize}

With these in mind, a hallucination map  $\cHMnull$ in the generalized null space $\Unull$ is defined as follows. 
\begin{definition}[\textit{Generalized null space hallucination map}]
    Consider a pixelwise indicator function $\mathds{1} : \mathbb{R}^N \rightarrow \mathbb{R}^N$ such that for any $\vartheta \in \mathbb{R}^N$,
    \begin{equation}
    [\mathds{1}(\vartheta)]_n = 
    \begin{cases}
    1,& \text{if  } [\vartheta]_n \neq 0\\
    0,& \text{if  } [\vartheta]_n  = 0.
    \end{cases}
    \end{equation}
    Then, the hallucination map $\coeff^{HM}_{null} \in \mathbb{E}^N$ can be defined as
    \begin{equation}\label{eq:hal_map_true_null}
    \hcoeff^{HM}_{null} \equiv  \mathds{1}(\hat{\coeff}_{null}) \odot (\hat{\coeff}_{null} - \coeff_{null}),
    \end{equation}
    where $\odot$ denotes the Hadamard product or element-wise multiplication.
    Note that the indicator function in the definition ensures that $\hcpinv$ does not possess any null space hallucinations, since no prior was imposed.
\end{definition}
It is important to highlight that, for the computation of the hallucination map in the generalized null space, one must have full knowledge of the generalized null component of the true object. This is in contrast to the hallucinations in the generalized measurement space, where the knowledge of the generalized measurement component of the true object  is not required.
This simply reflects that, according to the provided definitions, the generalized null space hallucination maps depict errors in the reconstructed null component of the object, while the generalized measurement space hallucination maps depict errors in the component of the object that can be stably reconstructed via a truncated pseudoinverse operator from the observed measurement data. 

This difference in the two definitions is associated with the fact that $\hcpinv$ is close to $\vec{H}^{+}_{P}\vec{H}\coeff$ if the measurement noise is small in the sense of Eq. \eqref{eq:stability}, and/or the model error is negligible. Hence, the proposed definition of $\hat{\coeff}^{HM}_{meas}$ is able to reveal the effect of the prior on the reconstructed generalized measurement space component, without requiring the true object. In this sense, there is no analog of a stably reconstructed component like $\hcpinv$ in the null space; hence invoking the true null component is necessary for defining $\hat{\coeff}^{HM}_{null}$. Note that due to our definition, $\hat{\coeff}^{HM}_{meas}$ may also be influenced by
the different noise propagation characteristics of the methods employed to form $\hcpinv$ and $\hat{\coeff}$ and therefore may not solely quantify errors associated with the prior. 

It should also be noted that the errors introduced by the prior in the measurement space can be remedied by adopting a reconstruction method that penalizes measurement space hallucinations without any prior knowledge of the object, e.g., via a data consistency constraint \cite{antun2020instabilities} or null space shuttle procedure \cite{deal1996nullspace}. Accordingly, for such constrained image
reconstruction methods, analyzing hallucinations in the null space is critical towards understanding the effect of the prior on the image estimate. 

Similar to the hallucination map in the generalized measurement space, the hallucination map $\faHMnull$ can be defined as
\color{black}
\begin{equation}\label{eq:hal_map_null_CD}
\faHMnull \equiv \sum_{n=1}^N [\hat{\theta}^{HM}_{null}]_n \expn.
\end{equation}

According to the proposed definitions, the truncated pseudoinverse solution $\hcpinv$ has zero hallucination in both the generalized measurement space and null space. However, that does not necessarily imply that $\hcpinv$ is without artifacts, since $\hcpinv$ ignores $\cnull$ completely. The computation of $\hcpinv$ leads to the recovery of only $\cmeas$ that can be estimated stably. When other regularized reconstruction methods attempt to reduce artifacts by imposing priors to estimate $\cnull$, a trade-off is made between the estimation of $\cmeas$ and $\cnull$ that can potentially lead to hallucinations in the generalized measurement space and null space.



\subsection{Specific hallucination maps}
\label{sec:specific}
The use of objective, or task-based, measures of image quality for evaluating imaging systems has been widely advocated \cite{barrett2013foundations}. However, the hallucination maps as defined in Section \ref{sec:definition} do not incorporate any task-specific information. In particular, $\hat{\coeff}^{HM}_{null}$ may contain an abundance of structures or textures, some of which may not confound an observer on a specified diagnostic task. Hence, it may be useful to identify those structures or textures in the hallucination maps that are task-relevant. 
One possible way to accomplish this is to process the hallucination map via an image processing transformation $T$, such that 
potentially task-relevant features or textures  are localized while others are suppressed \cite{castellano2004texture, chowdhary2020segmentation}.
Formally, this can be described as:
\begin{equation}\label{eq:shm_definition}
\hat{\coeff}^{SHM}_{null} = T\hat{\coeff}^{HM}_{null},
\end{equation}
where the processed pixel map $\hat{\coeff}^{SHM}_{null}$  that preserves task-specific information  is referred to as a \textit{specific hallucination map}. Note that the design of the transformation $T$ is application-dependent, as it should localize those structures or textures from the hallucination map that are relevant to a specified task.  Moreover, the specification of the observer (which could be a human or computational procedure) who will perform the task should also influence the design of $T$, as the extent to which hallucinations  impact observer performance will vary. While requiring significant effort to formulate, specific hallucination maps open up the possibility of comparing reconstruction methods based on their propensities for creating hallucinations that influence task-performance.

The complete procedure for computing measurement and null space hallucination maps, as well as the specific hallucination map, is presented in Algorithm \ref{alg}.

\floatname{algorithm}{Algorithm}
\begin{algorithm}[!t]
\caption{Procedure for computation of measurement and null space hallucination maps from measurement data $\vec{g}$, system matrix $\vec{H}$, true object $\coeff$ and reconstructed image $\hcoeff$}\label{alg}

\begin{algorithmic}[1]
\State Compute the truncated pseudoinverse solution:
\begin{equation}
    \hcpinv = \MP_P \vec{g}\nonumber.
\end{equation}

\State Compute the generalized measurement component of $\hcoeff$:
\begin{align}
    \hcmeas &= \mathcal{P}_{meas}\hcoeff = \MP_P \vec{H}\hcoeff.\nonumber
\end{align}

\State Compute the generalized null components of $\coeff$ and $\hcoeff$:
\begin{align}
    \cnull &= \mathcal{P}_{null}\coeff = [\vec{I}_N - \MP_P \vec{H}]\coeff,\nonumber\\
    \hcnull &= \mathcal{P}_{null}\hcoeff = [\vec{I}_N - \MP_P \vec{H}]\hcoeff.\nonumber
\end{align}

\State \textbf{Measurement space hallucination map}:
\begin{equation}
    \hcmeas^{HM} = \hcmeas-\hcpinv.\nonumber
\end{equation}

\State \textbf{Null space hallucination map}:
\begin{equation}
    \hcoeff^{HM}_{null} =  \mathds{1}(\hat{\coeff}_{null}) \odot (\hat{\coeff}_{null} - \coeff_{null})\nonumber.
\end{equation}

\State Apply image processing transformation $T$ on $\coeff^{HM}_{null}$ to obtain the specific hallucination map:
\begin{equation}
    \hat{\coeff}^{SHM}_{null} = T\hat{\coeff}^{HM}_{null}.\nonumber
\end{equation}
\end{algorithmic}
\end{algorithm}

\section{Numerical Studies}
\label{sec:Numerical studies}

\begin{figure*}[!htb]
    \centering
     \includegraphics[width=0.97\linewidth]{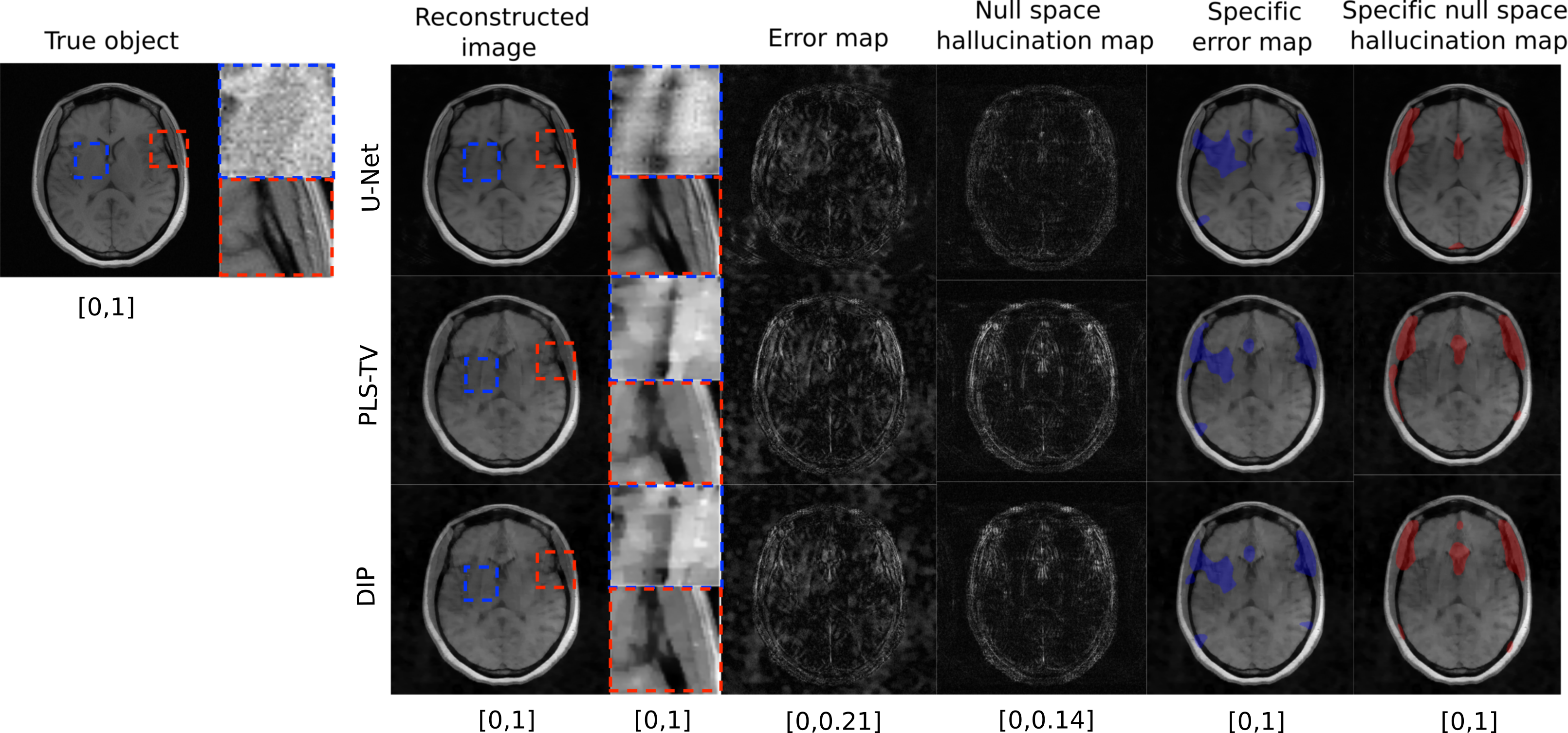}
      \caption{Example of a true object and reconstructed images along with error maps and hallucination maps (null space) for IND data with different reconstruction methods -- U-Net (top), PLS-TV (middle) and DIP (bottom). Expanded regions are shown to the right of the reconstructed images.
      The specific error map (blue) and specific null space hallucinations map (red) are overlaid on the reconstructed images for each method. The image estimated by the U-Net method has visibly lower hallucinations in the null space compared to PLS-TV and DIP. The region within the red bounding box is one of the locations that contains hallucinations for all the reconstruction methods. In this region, the U-Net method shows mild hallucinations compared to PLS-TV and DIP. Fine structures in this region appear to be oversmoothed in the image estimates obtained by use of PLS-TV and DIP.  A false structure is also shown (within the blue bounding box region) that appears for all the reconstruction methods due to the phase noise and not due to the imposed prior, and hence cannot be classified as a hallucination.
      \vspace{-0.5cm}}
      \label{fig:recons_ind}
\end{figure*}

Numerical studies were conducted to demonstrate the utility of the proposed hallucination maps.
Although the focus of these preliminary studies is on null space hallucination maps, the presented analyses could readily be repeated by use of measurement space hallucination maps. Hallucination maps were employed to compare the behavior of data-driven and model-based image reconstruction methods under different conditions.

\vspace{-0.3cm}


\subsection{Stylized imaging system}
\label{subsection:imaging system}
A stylized two-dimensional (2D) single-coil magnetic resonance (MR) imaging system was considered. It should be noted that the assumed imaging operator was not intended to accurately model a real-world MR imager. Instead, the purpose of the presented simulation studies is only to demonstrate the potential utility of hallucination maps. Hence physical factors such as coil sensitivity and bias field inhomogeneity were not considered. 
Fully-sampled k-space data were emulated by applying the 2D Fast Fourier Transform (FFT) on the digital objects described below.  Independent and identically distributed Gaussian noise  was added to the real and imaginary components of the complex-valued k-space data \cite{aja2016statistical} in the training dataset for the U-Net as well as in the test data during evaluation with different reconstruction methods. Additionally, in the test dataset, zero-mean random uniform phase noise \cite{xiaoyu2017compressed} was introduced into the k-space measurements to emulate modelling errors \cite{barrett2013foundations}.
A uniform Cartesian undersampling mask with an undersampling factor of 3 was applied on the fully-sampled k-space data to obtain undersampled measurements, as shown in Fig. S.1 in the Supplementary file. The k-space lines that were not sampled were subsequently zero-filled. The Moore-Penrose pseudoinverse $\MP$ was applied by performing the inverse 2D Fast Fourier Transform (IFFT) on the zero-filled k-space data. Since the true pseudoinverse was considered without any truncation of singular values, the hallucination map in the generalized null space in our studies corresponds to the hallucination  map in the true null space. 
\vspace{-0.2cm}
\subsection{Reconstruction methods}
Both data-driven and non-data-driven image reconstruction methods were investigated. The data-driven method considered was a U-Net based method \cite{jin2017deep, han2018framing, hyun2018deep}, 
which learns a mapping from an initial image estimate that contains artifacts due to undersampling to an accurate estimate of the true object.  In our studies, the initial image estimate that was input to the U-Net was obtained by applying the pseudoinverse on the k-space data. Two different non-data-driven reconstruction methods were considered. The first method, which is known as penalized least-squares with total variation (PLS-TV)\cite{sidky2008image}, involves solving a least-squares optimization problem with a total variation penalty \cite{sidky2008image}. The second  method  is known as deep image prior (DIP) \cite{ulyanov2018deep,van2018compressed}, where the reconstructed estimate is constrained to lie in the range of an untrained deep network  \cite{goodfellow2016deep} such that the estimate agrees with the observed measurements in a least-squares sense. These reconstruction methods are described in detail in Sec. S.I of the Supplementary file.
\vspace{-0.2cm}
\subsection{Training, validation and test data}
For the U-Net based reconstruction method, training was performed on 2D axial adult brain MRI images from the NYU fastMRI Initiative database \cite{zbontar2018fastmri}. These will be referred to as the in-distribution (IND) images.   The training and validation datasets contained 2500 and 500 images, respectively. 
For testing, both IND and out-of-distribution (OOD) images were considered. The OOD images were obtained from a pediatric epilepsy resection MRI dataset \cite{maallo2020effects}. Both the IND and OOD testing datasets contained 69 images. It should be noted that the OOD images differed from the IND images in several aspects, such as the nature of the objects (adult for IND and pediatric for OOD) as well as the use of different MR systems used to obtain the true object images in each case. All images were of dimension $320 \times 320$.


After creating the training, validation and test datasets, neural network training was performed with the IND training and validation datasets for the U-Net method. At test time, images were reconstructed from both IND and OOD test datasets using the U-Net, PLS-TV and DIP methods. Details regarding the implementation of these methods are presented in Sec. S.II of the Supplementary file.
\vspace{-0.2cm}
\subsection{Computation of hallucination maps}
After images were reconstructed  from the testing data, null space hallucination maps $\hat{\coeff}^{HM}_{null}$ were computed. The quantities  $\hat{\coeff}_{null}$ and ${\coeff}_{null}$, as required by Eq. \eqref{eq:hal_map_true_null}, were computed according to Eq. \eqref{eq:project_null_MP}. Subsequently, specific null space hallucination maps  $\hat{\coeff}^{SHM}_{null}$ were also computed. 
In this preliminary study, these maps were designed for the purpose of localizing regions where coherent structures, as opposed to random errors,  were present in $\hat{\coeff}^{HM}_{null}$.  Such structured hallucinations could be relevant to certain signal detection tasks.
To accomplish this, the transformation $T$ in Eq.\ (\ref{eq:shm_definition}) was implemented as follows.
\begin{figure}[!htb]
    \centering
     \includegraphics[width=\linewidth]{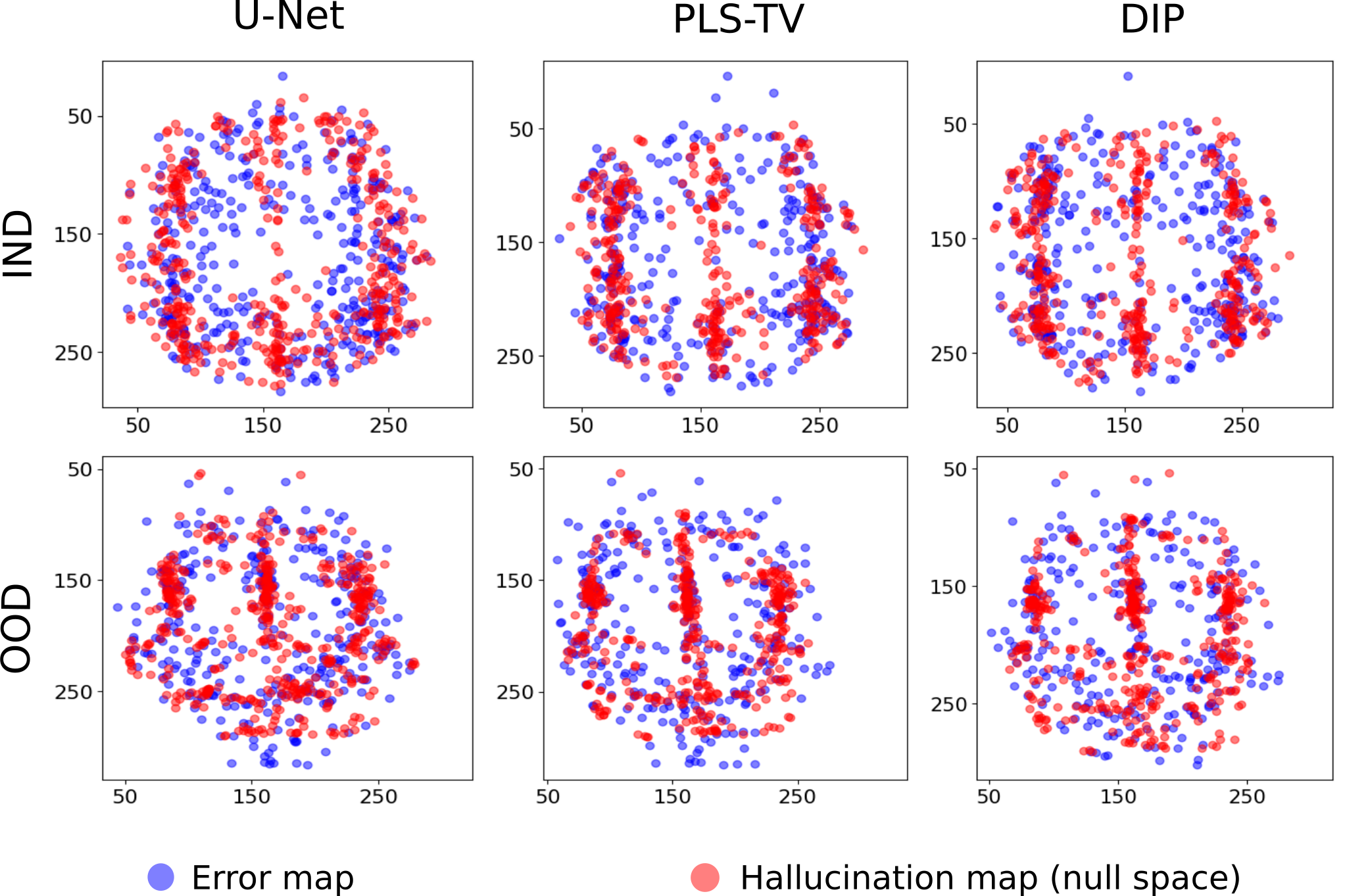}
      \caption{Scatter plots for centroids of localized regions in specific error maps and specific null space hallucination maps with different reconstruction methods for IND (top) and OOD (bottom) data. Note that for each type of data distribution and for all the reconstruction methods, the centroids of the regions detected from the error map have a higher variance compared to the hallucination map as well as some degree of non-overlap.}
      \label{fig:scatter_plots}
\end{figure}
First, the region of support of each object was identified using Otsu's method \cite{jain1989fundamentals} and binary support masks were formed for each object. The support masks were applied on the $\hat{\coeff}^{HM}_{null}$ such that errors in the reconstructed image that lie outside the region of support could be ignored. Subsequently, histogram equalization was performed. A 2D Gaussian filter with kernel width of 7 was applied on the histogram-equalized map in order to obtain a smooth distribution of intensities across the hallucination map. The width of the Gaussian filter was chosen heuristically in this study. Finally, a binary threshold was applied where the cut-off value was set to the 95-th percentile of intensity values in the processed map, such that intensities below the threshold were set to zero and intensities above the threshold were set to 1. 
From the thresholded maps, connected components that had a size of less than 100 pixels ($\approx 0.1\%$ of total number of pixels in each image) were eliminated to remove localized regions with negligible dimensions, resulting in the specific hallucination maps $\hat{\coeff}^{SHM}_{null}$ for our studies. This procedure for computing the action of $T$ was repeated for all $\hat{\coeff}^{HM}_{null}$ computed from both the IND and OOD test datasets for each reconstruction method.  It should be noted that this procedure  serves only as a simplistic example of the computation of a specific hallucination map, and there is no suggestion that it is optimal in any sense.

Finally, conventional error maps were computed as the difference between the reconstructed estimate $\hat{\coeff}$ and the true object $\coeff$. In order to demonstrate the potential utility of the specific hallucination maps over processed versions of conventional error maps, \emph{specific error maps} were formed by acting $T$ on the error maps. The codes employed in our numerical studies are available at \url{https://github.com/uiuc-comp-imaging-sci/hallucinations-tomo-recon}.


\begin{figure*}[!htb]
    \centering
     \includegraphics[width=0.97\linewidth]{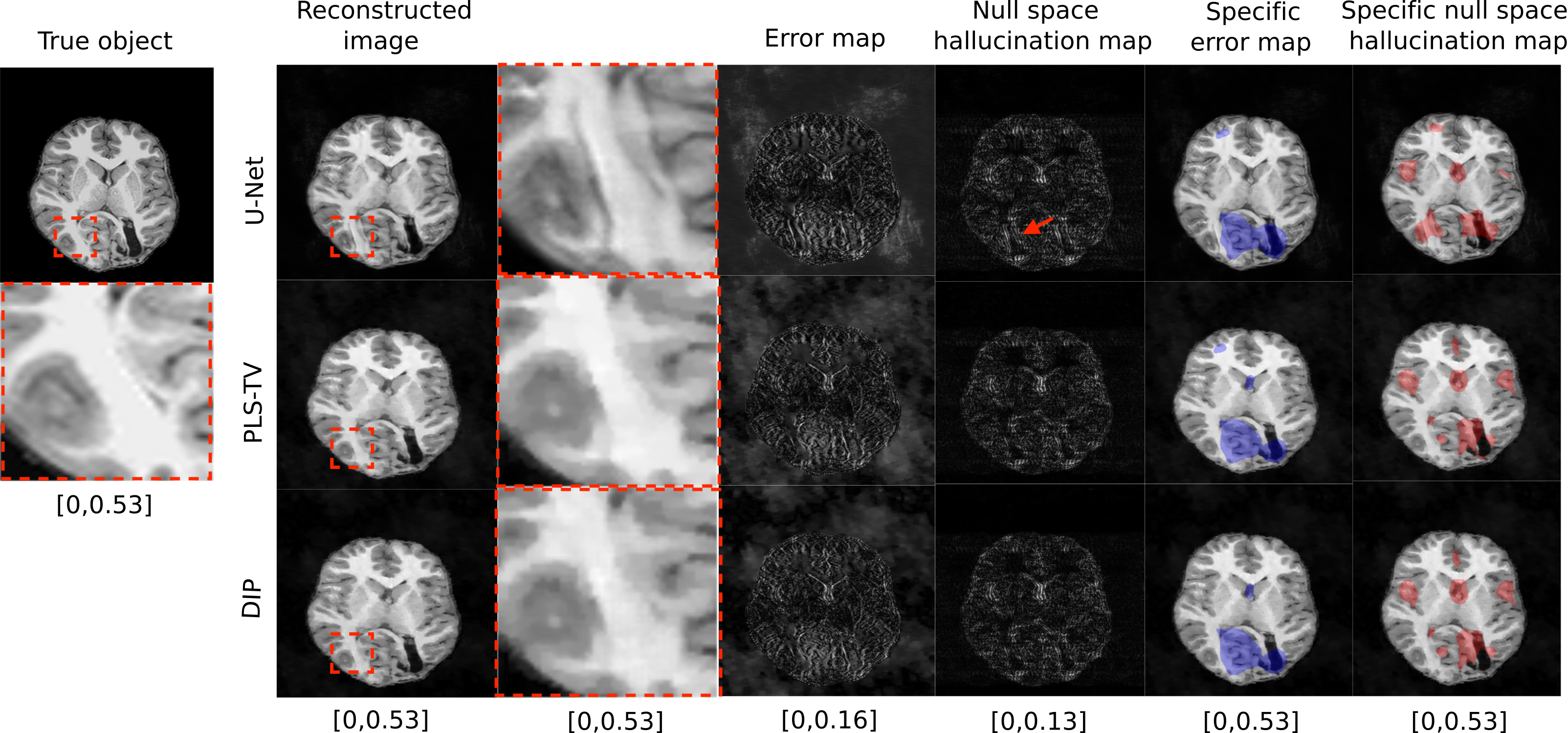}
      \caption{Example of true object and reconstructed images along with error map and hallucination maps (null space) for OOD data with different reconstruction methods -- U-Net (top), PLS-TV (middle) and DIP (bottom). 
      Expanded regions are shown to the right of the reconstructed images. The specific error map (blue) and specific null space hallucinations map (red) are overlaid on the reconstructed images for each method. The image estimated by the U-Net method has some distinct false structures (region within red bounding box) that do not exist in the reconstructed images obtained by using PLS-TV and DIP. This region is also highlighted in the specific null space hallucination map for the U-Net method which suggests that the false structure is a hallucination.}
      \vspace{-0.5cm}
      \label{fig:recons_ood}
\end{figure*}
\section{Results}
\label{sec:results}
The numerical results are organized as follows. First, an illustration of hallucination maps is provided for different reconstruction methods, in order to demonstrate their utility in highlighting false structures that may be introduced due to the imposed prior. Differences in the null space hallucination maps corresponding to the data-driven U-Net method when applied to IND and OOD data are examined. This is followed by a demonstration of the difference in the quantitative performance of the U-Net method on IND and OOD data. The performance of the U-Net is compared with the non-data driven methods in our studies -- PLS-TV and DIP -- in terms of metrics derived by use of null space hallucination maps. 

\subsection{Differences between error and hallucination maps}
\label{sec:differences}
Reconstructed images and corresponding error maps and null space hallucination maps from an IND measurement are shown in Fig. \ref{fig:recons_ind}. It can be observed that, for all the reconstruction methods, the error map and the null space hallucination map have different characteristics in some regions of the image. This is because the error map contains false structures due to hallucinations as well as all other factors, whereas the null space hallucination map only contains errors due to the imposed prior. These differences can also be observed from the computed specific error maps and specific null space hallucination maps.
As expected, the U-Net method performs well, leading to mostly low intensity regions in the null space hallucination map. In one of the regions that is featured in the specific hallucination map for all the reconstruction methods, it can be seen that the U-Net has lower hallucinations since it is able to faithfully recover fine structures in the region. Such fine structures were oversmoothed in the reconstructed images that were obtained by use of the PLS-TV and DIP methods, leading to higher hallucinations. On the other hand, all the reconstructed images also contain a distinct false structure that is revealed in the specific error map but not the specific hallucination map. This is an example of a false structure that can exist in reconstructed images, but may not necessarily be classified as a hallucination.

To further demonstrate the different characteristics of error maps and null space hallucination maps for this IND study,
scatter plots of the centroids of the detected regions in each type of map corresponding to the ensemble of IND reconstructed images from all three reconstruction methods are shown in Fig. \ref{fig:scatter_plots} (top row). From these scatter plots, it can be observed that there is a high amount of variance in the locations of the detected regions in the specific error maps as compared to the detected regions in the specific hallucination maps. The latter typically appear in similar regions across the ensemble of reconstructed images for all the methods. Furthermore, the concentrations of centroids for the detected regions in both types of maps have some degree of non-overlap. These observations reflect the fact that, due to additional sources of error such as measurement noise and model error that are also typically random in nature, the regions in the reconstructed images that are revealed by the error map can sometimes be different from those revealed by the null space hallucination map that considers error only due to an inaccurate prior.

As the distribution shifts to OOD, as shown in Fig. \ref{fig:recons_ood}, the null space hallucination map for the U-Net method appears comparable to the hallucination maps obtained by use of PLS-TV and DIP. False structures that can be identified as hallucinations appear in the image reconstructed by the U-Net method.
The higher error for the U-Net method is a result of the change of distribution and the method's inability to generalize well to data that are significantly out of distribution with respect to the training data. The change of distribution results in significant inaccuracies in the null component of the reconstructed estimate produced by the U-Net. Under such circumstances, it can be useful to identify and localize hallucinations due to inaccuracies in the imposed data-driven regularization through 
the null space hallucinations. 

As shown in Fig. \ref{fig:recons_ind} and consistent with the IND results discussed above,
the localized regions detected in the specific error map and specific hallucination map for the OOD cases are generally different.
Scatter plots of the centroids of the detected regions in the specific error maps and specific hallucination maps confirm this and are displayed in Fig. \ref{fig:scatter_plots} (bottom row).
For all the reconstruction methods, the error map centroids again have a higher variance and are located away from clusters of hallucination map centroids in some regions.
In other words, under such circumstances, one cannot rely on only the error maps without considering the corresponding hallucination maps in order to estimate where hallucinations due to the imposed prior are likely to be localized in a reconstructed image. 

Although hallucination maps can reveal false structures, the impact of the false structures on specific applications requires further analysis. For example, a false structure may be classified as a \textit{false positive structure} or a \textit{false negative structure}\cite{huang2020data,cheng2020addressing}. A false positive structure is one which is absent in the true object but present in the reconstructed image, whereas a false negative structure denotes the opposite. 
While an important topic, the classification of hallucinations is beyond the scope of this paper.

\subsection{Investigation of structured hallucinations}
\label{sec:recon_performance}
\begin{figure}[!htbp]
    \centering
     \includegraphics[width=\linewidth]{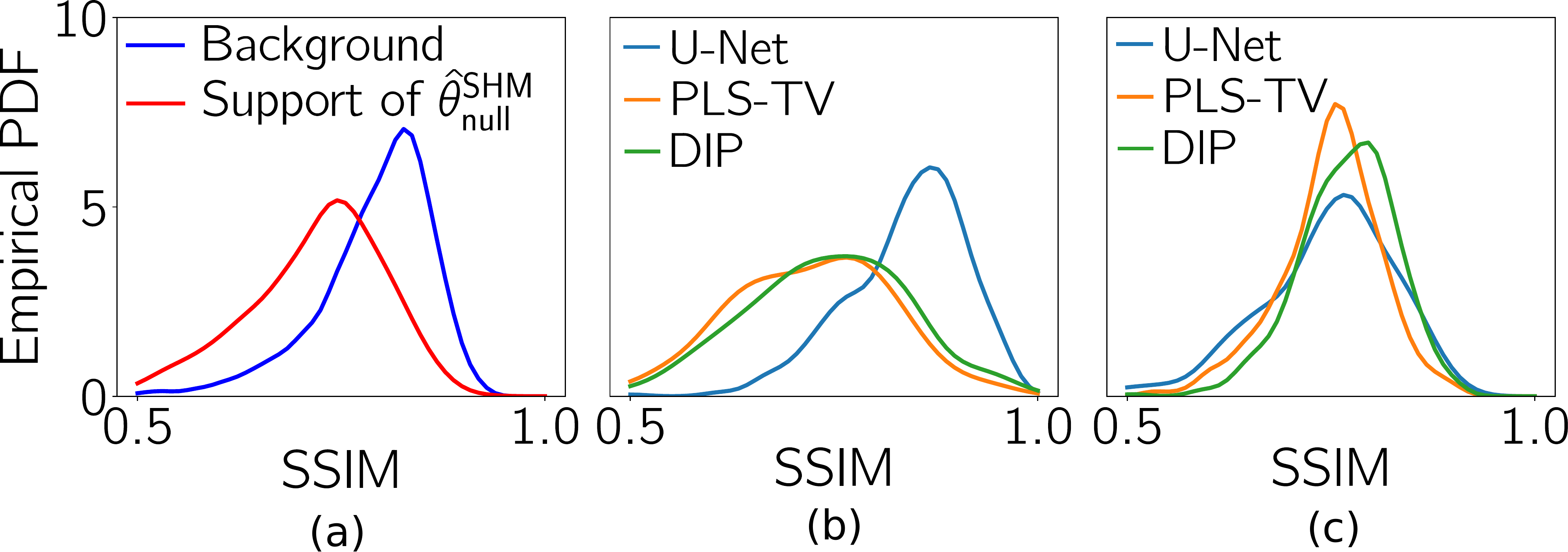}
      \caption{(a) Empirical PDF of SSIM values in the structured hallucination regions (support of $\hat{\coeff}^{SHM}_{null}$) and the regions spanned by the remaining pixels in the support of the image (background), respectively, for the U-Net method with OOD data. Empirical PDFs of SSIM values in the structured hallucination regions for all three reconstruction methods with (b) IND and (c) OOD data respectively.
      }
      \label{fig:compare_ssim}
\end{figure}

Additional studies were conducted to validate that the specific hallucination maps actually revealed regions in the image that contain significant errors.
To accomplish this, two empirical probability distribution functions (PDFs) were estimated that describe the  average SSIM values computed over two non-overlapping regions in the reconstructed images for the OOD case. One region corresponded to the support of the specific hallucination maps described above and the second region was spanned by all other pixels in the image.
The two empirical PDFs are shown in Fig. \ref{fig:compare_ssim}a and reveal that
the mode of the  distribution corresponding to the SSIM averaged over the structured hallucination regions is demonstrably lower than that describing the average SSIM values over the background regions.  


The empirical PDFs that described  the SSIM value averaged over the structured hallucination regions were also compared for each of the three reconstruction methods.
As shown in Fig. \ref{fig:compare_ssim}b, for the IND case, the images reconstructed by use of the U-Net  had significantly higher SSIM values, on average, in the structured hallucination regions as compared to both the PLS-TV and DIP methods.  This can be attributed to network training with a sufficiently large amount of IND data. 
However, for the OOD case in Fig. \ref{fig:compare_ssim}c, because null space hallucinations increased for the U-Net method, the corresponding reconstructed images had lower SSIM values on average as compared with DIP in the support of the null space hallucination maps. The medians of ensemble SSIM values in these support regions for all the reconstruction methods with IND and OOD data are shown in Table \ref{tab:ssim}.
It should be noted that, for both the IND and OOD cases, the DIP method was implemented with the same network architecture as the U-Net based  method. Thus, when there is a shift in the testing data distribution, some data-driven methods such as the U-Net method may not provide any significant improvement in the estimate of the null component compared to model-based methods that do not employ training data. However, the data-driven methods involve the additional risk of hallucinating false structures.
These observations gained through hallucination maps provide insight into the impact of the data-driven nature of the prior imposed by pre-trained neural networks.

\subsection{Bias maps and hallucinations}
\begin{figure}[!htbp]
    \centering
     \includegraphics[width=\linewidth]{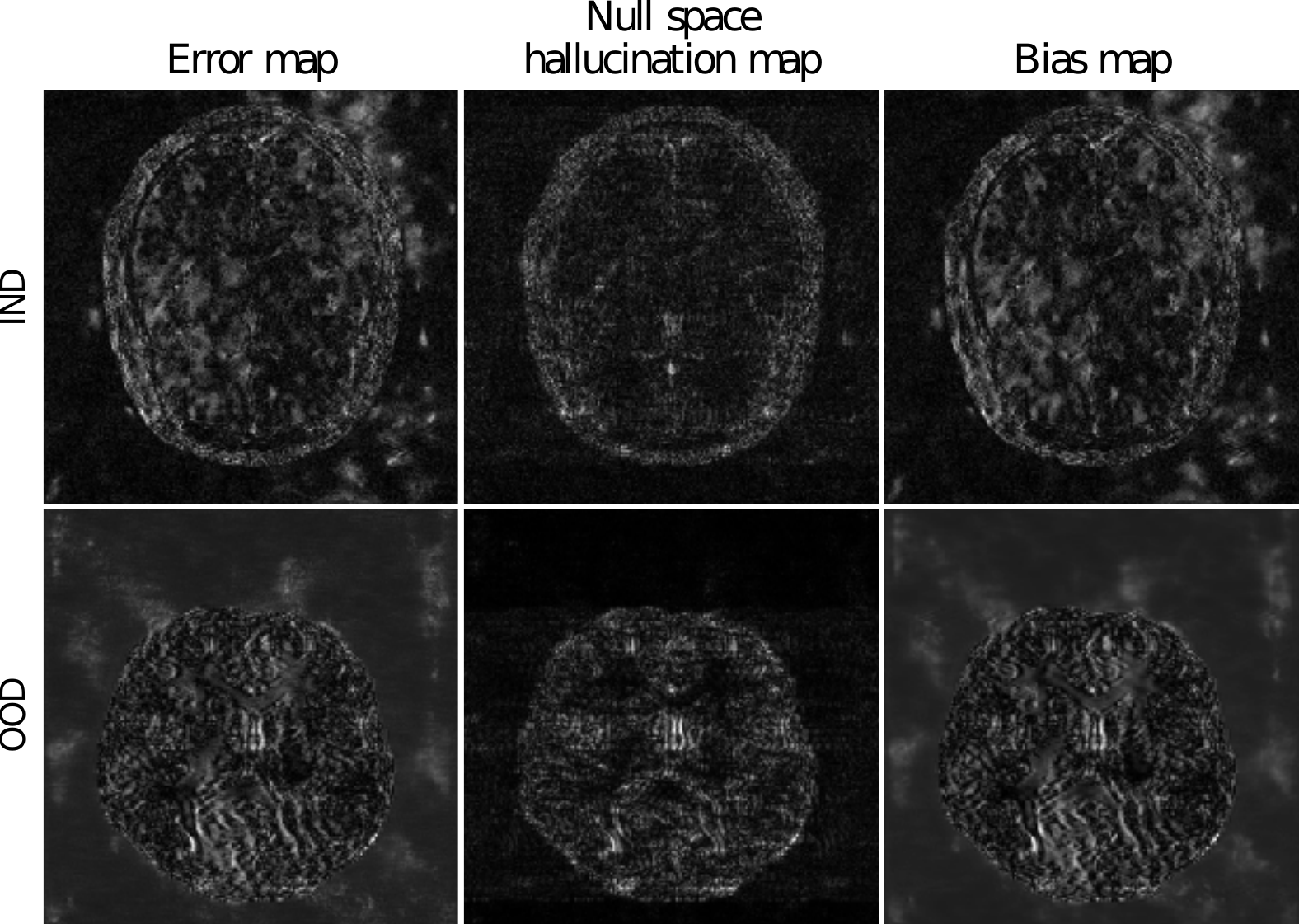}
      \caption{An error map, a null space hallucination map and a bias map for IND and OOD images estimated by use of the U-Net method. The corresponding true objects are shown in Figures \ref{fig:recons_ind} and \ref{fig:motivating_image}b respectively. The bias map was computed over a dataset of 100 images estimated from a single set of simulated measurements with fixed phase noise and different realizations of the iid Gaussian noise. The bias map contains contributions from both the model error, as well as inaccuracies in the prior.}
      \label{fig:bias_maps}
\end{figure}

A \textit{bias map}, defined as 
\begin{align}
    \vec{b} \equiv \mathbb{E}\hat{\coeff} - \coeff,
\end{align}
determines the expected deviation of an image estimate from the true object, and as such, may include contributions from an incorrect prior, as well as those from incorrect measurement and noise models. Hence, the bias map may be correlated with the hallucination maps, but may display significant differences from it based on the average behavior of the inaccuracies in the measurement and noise models. For example, Fig. \ref{fig:bias_maps} shows the bias map computed using a dataset of images estimated from simulated undersampled MRI measurements with fixed phase noise and iid Gaussian additive noise, along with the error map and the null space hallucination map for an IND and an OOD image. The corresponding true objects are shown in Figures \ref{fig:recons_ind} and \ref{fig:motivating_image}b respectively. Figure \ref{fig:bias_maps} shows that the bias map retains clusters of artifacts from the error map that are due to the phase noise. Hence, although the bias maps are correlated with both the hallucination maps and the error map, each provides a different kind of information. 
\color{black}


\vspace{0.3cm}

\begin{table}[!htbp]
    \centering
    \begin{tabular}{c|ccc}
    \toprule
     Data distribution    &  U-Net & PLS-TV & DIP\\
     \midrule
     IND             & \textbf{0.84}   & 0.71   & 0.73\\
     OOD             & 0.75   & 0.73   & \textbf{0.76}\\
     \bottomrule
    \end{tabular}
    \caption{Median of SSIM values from ensembles of images reconstructed by use of the U-Net, PLS-TV and DIP methods that were computed in the support region of specific null space hallucination maps. In these regions, the U-Net method has the highest median SSIM for IND data, while for OOD data the DIP method has the highest median SSIM.}
    \label{tab:ssim}
\end{table}

\section{Summary and Conclusion}
\label{sec:discussion}


While regularization via sparsity-promoting penalties in an optimization-based reconstruction framework is commonly employed, emerging learning-based methods that employ deep neural networks have shown the potential to improve reconstructed image quality further by learning priors from existing data.  However, an analysis of the prior information learned by deep networks and their ability to generalize to data that may lie outside the training distribution is still being explored. Additionally, there are open questions and concerns about the stability of such networks when applied for image reconstruction. While it has been understood that use of an inaccurate prior might lead to false structures, or hallucinations, being introduced in the reconstructed image,   formal definitions for hallucinations within the context of tomographic image reconstruction have not been reported.

In this work, by use of concepts from linear operator theory, formal definitions for hallucination maps in linear tomographic imaging problems are introduced. These provide the opportunity to isolate and visualize image hallucinations that are contained within the measurement or null spaces of a linear imaging operator.  The measurement space hallucination map permits the analysis of errors in the measurement space component of a reconstructed object estimate with respect to the component of the object that can be stably computed from a given set of measurement data.  Alternatively, the null space hallucination map permits analysis of errors in the null space component of a reconstructed object estimate with respect to the true object null space component. These errors are caused solely by the reconstruction prior. Both maps can be employed to systematically investigate the impact of different priors utilized in image reconstruction methods.  Finally, the notion of a specific hallucination map was also introduced, which can be formulated to reveal hallucinations that are relevant to a specified image-based inference.

Numerical studies were performed with simulated undersampled measurements from a stylized single-coil MRI system. Both data-driven and non-data-driven methods were investigated to demonstrate the utility of the proposed hallucination maps. It was observed that null space hallucination maps can be particularly useful as compared to traditional error maps when assessing the effect of data-driven regularization strategies with out-of-distribution data. Furthermore, it was shown that structured hallucinations with data-driven methods that are caused due to a shift in the data distribution may ultimately lead to significant artifacts in the reconstructed image. 


The computation of the projection operations as described in Eq. \eqref{eq:project_meas_MP} and Eq. \eqref{eq:project_null_MP} via the SVD may be infeasible for large-scale problems. 
Wilson and Barrett \cite{wilson1998decomposition} proposed an iterative method to compute $\cmeas$ and $\cnull$ without explicit computation of the SVD of $\vec{H}$. Alternatively, randomized SVD  \cite{halko2011finding} is a relatively computationally efficient algorithm that can be employed to estimate these quantities. Kuo \textit{et al.}\cite{kuo2021learning} recently proposed a method to learn null space projection operations that can significantly reduce the computational burden. It may also be expected that the importance of analyzing hallucinations in image reconstruction can further stimulate the development of efficient methods for implementing projection operators. The development of such computationally efficient methods for large-scale problems remains an active area of research. 

It should be noted that the proposed definition of hallucination maps is general and can be applied to any linear imaging system and reconstruction method, provided that the computation of the projection operators $\mathcal{P}_{meas}$ and $\mathcal{P}_{null}$ is feasible. Depending on the sampling pattern involved in the data acquisition process, different system matrices $\vec{H}$ will have different null space characteristics. This, in turn, may lead to different properties in the corresponding hallucination maps that would allow a comparison of reconstruction methods under a variety of data acquisition strategies.

The proposed framework is most useful in situations where the generalized null component of the true object is significant and hence strong priors need to be incorporated in the reconstruction method via regularization. If the generalized null component is relatively small compared to the generalized measurement component, the need for strong regularization during reconstruction is diminished. This, in turn, would imply that hallucinations are likely to be minimal or non-existent due to the imposed weak regularization and hence computing hallucination maps may not be necessary. In such situations, computing only the error map may be sufficient to assess the reconstruction method.

There remain important topics for future investigation. 
Beyond the framework presented, it will be important to derive objective figures-of-merit (FOMs) from ensembles of hallucination maps. Furthermore, the probability of occurrence of hallucinations can be potentially quantified from ensembles of hallucination maps.  
While understanding the interplay between hallucinations and image reconstruction priors is important in preliminary studies,
ultimately, image reconstruction methods should be objectively evaluated with consideration of all physical and statistical factors.

\vspace{-0.2cm}

\bibliography{DLRHM}{}

\begin{thebibliography}{10}
\providecommand{\url}[1]{#1}
\csname url@samestyle\endcsname
\providecommand{\newblock}{\relax}
\providecommand{\bibinfo}[2]{#2}
\providecommand{\BIBentrySTDinterwordspacing}{\spaceskip=0pt\relax}
\providecommand{\BIBentryALTinterwordstretchfactor}{4}
\providecommand{\BIBentryALTinterwordspacing}{\spaceskip=\fontdimen2\font plus
\BIBentryALTinterwordstretchfactor\fontdimen3\font minus
  \fontdimen4\font\relax}
\providecommand{\BIBforeignlanguage}[2]{{%
\expandafter\ifx\csname l@#1\endcsname\relax
\typeout{** WARNING: IEEEtran.bst: No hyphenation pattern has been}%
\typeout{** loaded for the language `#1'. Using the pattern for}%
\typeout{** the default language instead.}%
\else
\language=\csname l@#1\endcsname
\fi
#2}}
\providecommand{\BIBdecl}{\relax}
\BIBdecl

\bibitem{kak2002principles}
A.~C. Kak, M.~Slaney, and G.~Wang, ``Principles of computerized tomographic
  imaging,'' \emph{Medical Physics}, vol.~29, no.~1, pp. 107--107, 2002.

\bibitem{yang2016sparse}
A.~C.-Y. Yang, M.~Kretzler, S.~Sudarski, V.~Gulani, and N.~Seiberlich, ``Sparse
  reconstruction techniques in {MRI}: methods, applications, and challenges to
  clinical adoption,'' \emph{Investigative radiology}, vol.~51, no.~6, p. 349,
  2016.

\bibitem{donoho2006most}
D.~L. Donoho, ``{For most large underdetermined systems of linear equations the
  minimal $\ell_1$-norm solution is also the sparsest solution},''
  \emph{Communications on Pure and Applied Mathematics: A Journal Issued by the
  Courant Institute of Mathematical Sciences}, vol.~59, no.~6, pp. 797--829,
  2006.

\bibitem{donoho2003optimally}
D.~L. Donoho and M.~Elad, ``Optimally sparse representation in general
  (nonorthogonal) dictionaries via $\ell_1$ minimization,'' \emph{Proceedings
  of the National Academy of Sciences}, vol. 100, no.~5, pp. 2197--2202, 2003.

\bibitem{candes2006stable}
E.~J. Candes, J.~K. Romberg, and T.~Tao, ``Stable signal recovery from
  incomplete and inaccurate measurements,'' \emph{Communications on Pure and
  Applied Mathematics: A Journal Issued by the Courant Institute of
  Mathematical Sciences}, vol.~59, no.~8, pp. 1207--1223, 2006.

\bibitem{sidky2008image}
E.~Y. Sidky and X.~Pan, ``Image reconstruction in circular cone-beam computed
  tomography by constrained, total-variation minimization,'' \emph{Physics in
  Medicine \& Biology}, vol.~53, no.~17, p. 4777, 2008.

\bibitem{wang2016perspective}
G.~Wang, ``A perspective on deep imaging,'' \emph{IEEE access}, vol.~4, pp.
  8914--8924, 2016.

\bibitem{mccann2017convolutional}
M.~T. McCann, K.~H. Jin, and M.~Unser, ``Convolutional neural networks for
  inverse problems in imaging: A review,'' \emph{IEEE Signal Processing
  Magazine}, vol.~34, no.~6, pp. 85--95, 2017.

\bibitem{ravishankar2019image}
S.~Ravishankar, J.~C. Ye, and J.~A. Fessler, ``Image reconstruction: From
  sparsity to data-adaptive methods and machine learning,'' \emph{Proceedings
  of the IEEE}, vol. 108, no.~1, pp. 86--109, 2019.

\bibitem{huang2018some}
Y.~Huang, T.~W{\"u}rfl, K.~Breininger, L.~Liu, G.~Lauritsch, and A.~Maier,
  ``Some investigations on robustness of deep learning in limited angle
  tomography,'' in \emph{International Conference on Medical Image Computing
  and Computer-Assisted Intervention}.\hskip 1em plus 0.5em minus 0.4em\relax
  Springer, 2018, pp. 145--153.

\bibitem{gottschling2020troublesome}
N.~M. Gottschling, V.~Antun, B.~Adcock, and A.~C. Hansen, ``The troublesome
  kernel: why deep learning for inverse problems is typically unstable,''
  \emph{arXiv preprint arXiv:2001.01258}, 2020.

\bibitem{antun2020instabilities}
V.~Antun, F.~Renna, C.~Poon, B.~Adcock, and A.~C. Hansen, ``On instabilities of
  deep learning in image reconstruction and the potential costs of {AI},''
  \emph{Proceedings of the National Academy of Sciences}, 2020.

\bibitem{laves2020uncertainty}
M.-H. Laves, M.~T{\"o}lle, and T.~Ortmaier, ``Uncertainty estimation in medical
  image denoising with bayesian deep image prior,'' in \emph{Uncertainty for
  Safe Utilization of Machine Learning in Medical Imaging, and Graphs in
  Biomedical Image Analysis}.\hskip 1em plus 0.5em minus 0.4em\relax Springer,
  2020, pp. 81--96.

\bibitem{asim2020invertible}
M.~Asim, M.~Daniels, O.~Leong, A.~Ahmed, and P.~Hand, ``Invertible generative
  models for inverse problems: mitigating representation error and dataset
  bias,'' in \emph{International Conference on Machine Learning}.\hskip 1em
  plus 0.5em minus 0.4em\relax PMLR, 2020, pp. 399--409.

\bibitem{kelkar2021compressible}
V.~A. Kelkar, S.~Bhadra, and M.~A. Anastasio, ``Compressible latent-space
  invertible networks for generative model-constrained image reconstruction,''
  \emph{IEEE Transactions on Computational Imaging}, 2021.

\bibitem{baker2002limits}
S.~Baker and T.~Kanade, ``Limits on super-resolution and how to break them,''
  \emph{IEEE Transactions on Pattern Analysis and Machine Intelligence},
  vol.~24, no.~9, pp. 1167--1183, 2002.

\bibitem{liu2005hallucinating}
W.~Liu, D.~Lin, and X.~Tang, ``Hallucinating faces: Tensorpatch
  super-resolution and coupled residue compensation,'' in \emph{2005 IEEE
  Computer Society Conference on Computer Vision and Pattern Recognition
  (CVPR'05)}, vol.~2.\hskip 1em plus 0.5em minus 0.4em\relax IEEE, 2005, pp.
  478--484.

\bibitem{wang2014comprehensive}
N.~Wang, D.~Tao, X.~Gao, X.~Li, and J.~Li, ``A comprehensive survey to face
  hallucination,'' \emph{International journal of computer vision}, vol. 106,
  no.~1, pp. 9--30, 2014.

\bibitem{fawzi2016image}
A.~Fawzi, H.~Samulowitz, D.~Turaga, and P.~Frossard, ``Image inpainting through
  neural networks hallucinations,'' in \emph{2016 IEEE 12th Image, Video, and
  Multidimensional Signal Processing Workshop (IVMSP)}.\hskip 1em plus 0.5em
  minus 0.4em\relax Ieee, 2016, pp. 1--5.

\bibitem{barrett2013foundations}
H.~H. Barrett and K.~J. Myers, \emph{Foundations of {I}mage {S}cience}.\hskip
  1em plus 0.5em minus 0.4em\relax John Wiley \& Sons, 2013.

\bibitem{anastasio2003basic}
M.~A. Anastasio and R.~W. Schoonover, ``Basic principles of inverse problems
  for optical scientists,'' \emph{digital Encyclopedia of Applied Physics}, pp.
  1--24, 2003.

\bibitem{bal2012introduction}
G.~Bal, ``Introduction to inverse problems,'' \emph{Lecture Notes-Department of
  Applied Physics and Applied Mathematics, Columbia University, New York},
  2012.

\bibitem{deal1996nullspace}
M.~M. Deal and G.~Nolet, ``Nullspace shuttles,'' \emph{Geophysical Journal
  International}, vol. 124, no.~2, pp. 372--380, 1996.

\bibitem{ravishankar2010mr}
S.~Ravishankar and Y.~Bresler, ``{MR} image reconstruction from highly
  undersampled k-space data by dictionary learning,'' \emph{IEEE transactions
  on medical imaging}, vol.~30, no.~5, pp. 1028--1041, 2010.

\bibitem{tiwari2019study}
S.~Tiwari, K.~Kaur, and K.~Arya, ``A study on dictionary learning based image
  reconstruction techniques for big medical data,'' in \emph{Handbook of
  Multimedia Information Security: Techniques and Applications}.\hskip 1em plus
  0.5em minus 0.4em\relax Springer, 2019, pp. 377--393.

\bibitem{ravishankar2012learning}
S.~Ravishankar and Y.~Bresler, ``Learning sparsifying transforms,'' \emph{IEEE
  Transactions on Signal Processing}, vol.~61, no.~5, pp. 1072--1086, 2012.

\bibitem{ravishankar2016data}
{S. Ravishankar and Y. Bresler}, ``{Data-driven learning of a union of
  sparsifying transforms model for blind compressed sensing},'' \emph{IEEE
  Transactions on Computational Imaging}, vol.~2, no.~3, pp. 294--309, 2016.

\bibitem{goodfellow2016deep}
I.~Goodfellow, Y.~Bengio, A.~Courville, and Y.~Bengio, \emph{Deep
  learning}.\hskip 1em plus 0.5em minus 0.4em\relax MIT press Cambridge, 2016,
  vol.~1.

\bibitem{mccann2019biomedical}
M.~T. McCann and M.~Unser, ``Biomedical image reconstruction: From the
  foundations to deep neural networks,'' \emph{arXiv preprint
  arXiv:1901.03565}, 2019.

\bibitem{hammernik2020machine}
K.~Hammernik and F.~Knoll, ``Machine learning for image reconstruction,'' in
  \emph{Handbook of Medical Image Computing and Computer Assisted
  Intervention}.\hskip 1em plus 0.5em minus 0.4em\relax Elsevier, 2020, pp.
  25--64.

\bibitem{stayman2012information}
J.~W. Stayman, J.~L. Prince, and J.~H. Siewerdsen, ``Information propagation in
  prior-image-based reconstruction,'' in \emph{Conference
  proceedings/International Conference on Image Formation in X-Ray Computed
  Tomography. International Conference on Image Formation in X-Ray Computed
  Tomography}, vol. 2012.\hskip 1em plus 0.5em minus 0.4em\relax NIH Public
  Access, 2012, p. 334.

\bibitem{castellano2004texture}
G.~Castellano, L.~Bonilha, L.~Li, and F.~Cendes, ``Texture analysis of medical
  images,'' \emph{Clinical radiology}, vol.~59, no.~12, pp. 1061--1069, 2004.

\bibitem{chowdhary2020segmentation}
C.~L. Chowdhary and D.~Acharjya, ``Segmentation and feature extraction in
  medical imaging: a systematic review,'' \emph{Procedia Computer Science},
  vol. 167, pp. 26--36, 2020.

\bibitem{aja2016statistical}
S.~Aja-Fern{\'a}ndez and G.~Vegas-S{\'a}nchez-Ferrero, ``Statistical analysis
  of noise in {MRI},'' \emph{Switzerland: Springer International Publishing},
  2016.

\bibitem{xiaoyu2017compressed}
F.~Xiaoyu, L.~Qiusheng, and S.~Baoshun, ``Compressed sensing mri with phase
  noise disturbance based on adaptive tight frame and total variation,''
  \emph{IEEE Access}, vol.~5, pp. 19\,311--19\,321, 2017.

\bibitem{jin2017deep}
K.~H. Jin, M.~T. McCann, E.~Froustey, and M.~Unser, ``Deep convolutional neural
  network for inverse problems in imaging,'' \emph{IEEE Transactions on Image
  Processing}, vol.~26, no.~9, pp. 4509--4522, 2017.

\bibitem{han2018framing}
Y.~Han and J.~C. Ye, ``Framing {U-Net} via deep convolutional framelets:
  Application to sparse-view {CT},'' \emph{IEEE transactions on medical
  imaging}, vol.~37, no.~6, pp. 1418--1429, 2018.

\bibitem{hyun2018deep}
C.~M. Hyun, H.~P. Kim, S.~M. Lee, S.~Lee, and J.~K. Seo, ``Deep learning for
  undersampled {MRI} reconstruction,'' \emph{Physics in Medicine \& Biology},
  vol.~63, no.~13, p. 135007, 2018.

\bibitem{ulyanov2018deep}
D.~Ulyanov, A.~Vedaldi, and V.~Lempitsky, ``Deep image prior,'' in
  \emph{Proceedings of the IEEE Conference on Computer Vision and Pattern
  Recognition}, 2018, pp. 9446--9454.

\bibitem{van2018compressed}
D.~Van~Veen, A.~Jalal, M.~Soltanolkotabi, E.~Price, S.~Vishwanath, and A.~G.
  Dimakis, ``Compressed sensing with deep image prior and learned
  regularization,'' \emph{arXiv preprint arXiv:1806.06438}, 2018.

\bibitem{zbontar2018fastmri}
J.~Zbontar, F.~Knoll, A.~Sriram, M.~J. Muckley, M.~Bruno, A.~Defazio,
  M.~Parente, K.~J. Geras, J.~Katsnelson, H.~Chandarana \emph{et~al.},
  ``fast{MRI}: An open dataset and benchmarks for accelerated {MRI},''
  \emph{arXiv preprint arXiv:1811.08839}, 2018.

\bibitem{maallo2020effects}
A.~M.~S. Maallo, E.~Freud, T.~T. Liu, C.~Patterson, and M.~Behrmann, ``Effects
  of unilateral cortical resection of the visual cortex on bilateral human
  white matter,'' \emph{NeuroImage}, vol. 207, p. 116345, 2020.

\bibitem{jain1989fundamentals}
A.~K. Jain, \emph{Fundamentals of digital image processing}.\hskip 1em plus
  0.5em minus 0.4em\relax Prentice-Hall, Inc., 1989.

\bibitem{huang2020data}
Y.~Huang, A.~Preuhs, M.~Manhart, G.~Lauritsch, and A.~Maier, ``{Data consistent
  CT reconstruction from insufficient data with learned prior images},''
  \emph{arXiv preprint arXiv:2005.10034}, 2020.

\bibitem{cheng2020addressing}
K.~Cheng, F.~Caliv{\'a}, R.~Shah, M.~Han, S.~Majumdar, and V.~Pedoia,
  ``{Addressing the false negative problem of deep learning MRI reconstruction
  models by adversarial attacks and robust training},'' in \emph{Medical
  Imaging with Deep Learning}.\hskip 1em plus 0.5em minus 0.4em\relax PMLR,
  2020, pp. 121--135.

\bibitem{wilson1998decomposition}
D.~W. Wilson and H.~H. Barrett, ``Decomposition of images and objects into
  measurement and null components,'' \emph{Optics Express}, vol.~2, no.~6, pp.
  254--260, 1998.

\bibitem{halko2011finding}
N.~Halko, P.-G. Martinsson, and J.~A. Tropp, ``Finding structure with
  randomness: Probabilistic algorithms for constructing approximate matrix
  decompositions,'' \emph{SIAM review}, vol.~53, no.~2, pp. 217--288, 2011.

\bibitem{kuo2021learning}
J.~Kuo, J.~Granstedt, U.~Villa, and M.~A. Anastasio, ``Learning a projection
  operator onto the null space of a linear imaging operator,'' in \emph{Medical
  Imaging 2021: Physics of Medical Imaging}, vol. 11595.\hskip 1em plus 0.5em
  minus 0.4em\relax International Society for Optics and Photonics, 2021, p.
  115953X.

\end{thebibliography}


\begin{thebibliography}{10}
\providecommand{\url}[1]{#1}
\csname url@samestyle\endcsname
\providecommand{\newblock}{\relax}
\providecommand{\bibinfo}[2]{#2}
\providecommand{\BIBentrySTDinterwordspacing}{\spaceskip=0pt\relax}
\providecommand{\BIBentryALTinterwordstretchfactor}{4}
\providecommand{\BIBentryALTinterwordspacing}{\spaceskip=\fontdimen2\font plus
\BIBentryALTinterwordstretchfactor\fontdimen3\font minus
  \fontdimen4\font\relax}
\providecommand{\BIBforeignlanguage}[2]{{%
\expandafter\ifx\csname l@#1\endcsname\relax
\typeout{** WARNING: IEEEtran.bst: No hyphenation pattern has been}%
\typeout{** loaded for the language `#1'. Using the pattern for}%
\typeout{** the default language instead.}%
\else
\language=\csname l@#1\endcsname
\fi
#2}}
\providecommand{\BIBdecl}{\relax}
\BIBdecl

\bibitem{chambolle2004algorithm}
A.~Chambolle, ``An algorithm for total variation minimization and
  applications,'' \emph{Journal of Mathematical Imaging and Vision}, vol.~20,
  no. 1-2, pp. 89--97, 2004.

\bibitem{beck2009fast}
A.~Beck and M.~Teboulle, ``Fast gradient-based algorithms for constrained total
  variation image denoising and deblurring problems,'' \emph{IEEE Transactions
  on Image Processing}, vol.~18, no.~11, pp. 2419--2434, 2009.

\bibitem{afonso2010augmented}
M.~V. Afonso, J.~M. Bioucas-Dias, and M.~A. Figueiredo, ``An augmented
  lagrangian approach to the constrained optimization formulation of imaging
  inverse problems,'' \emph{IEEE Transactions on Image Processing}, vol.~20,
  no.~3, pp. 681--695, 2010.

\bibitem{uecker2015berkeley}
M.~Uecker, F.~Ong, J.~I. Tamir, D.~Bahri, P.~Virtue, J.~Y. Cheng, T.~Zhang, and
  M.~Lustig, ``Berkeley advanced reconstruction toolbox,'' in \emph{Proc. Intl.
  Soc. Mag. Reson. Med}, vol.~23, no. 2486, 2015.

\bibitem{BART}
\BIBentryALTinterwordspacing
M.~U. \textit{et al.}, ``Bart toolbox for computational magnetic resonance
  imaging, doi: 10.5281/zenodo.592960,'' 2015. [Online]. Available:
  \url{http://mrirecon.github.io/bart/}
\BIBentrySTDinterwordspacing

\bibitem{jin2017deep}
K.~H. Jin, M.~T. McCann, E.~Froustey, and M.~Unser, ``Deep convolutional neural
  network for inverse problems in imaging,'' \emph{IEEE Transactions on Image
  Processing}, vol.~26, no.~9, pp. 4509--4522, 2017.

\bibitem{han2018framing}
Y.~Han and J.~C. Ye, ``Framing {U-Net} via deep convolutional framelets:
  Application to sparse-view {CT},'' \emph{IEEE transactions on medical
  imaging}, vol.~37, no.~6, pp. 1418--1429, 2018.

\bibitem{hyun2018deep}
C.~M. Hyun, H.~P. Kim, S.~M. Lee, S.~Lee, and J.~K. Seo, ``Deep learning for
  undersampled {MRI} reconstruction,'' \emph{Physics in Medicine \& Biology},
  vol.~63, no.~13, p. 135007, 2018.

\bibitem{antholzer2019deep}
S.~Antholzer, M.~Haltmeier, and J.~Schwab, ``Deep learning for photoacoustic
  tomography from sparse data,'' \emph{Inverse problems in science and
  engineering}, vol.~27, no.~7, pp. 987--1005, 2019.

\bibitem{ronneberger2015unet}
O.~Ronneberger, P.~Fischer, and T.~Brox, ``U-{N}et: Convolutional networks for
  biomedical image segmentation,'' in \emph{Medical Image Computing and
  Computer-Assisted Intervention -- MICCAI 2015}, N.~Navab, J.~Hornegger, W.~M.
  Wells, and A.~F. Frangi, Eds.\hskip 1em plus 0.5em minus 0.4em\relax Cham:
  Springer International Publishing, 2015, pp. 234--241.

\bibitem{drozdzal2016importance}
M.~Drozdzal, E.~Vorontsov, G.~Chartrand, S.~Kadoury, and C.~Pal, ``The
  importance of skip connections in biomedical image segmentation,'' in
  \emph{Deep Learning and Data Labeling for Medical Applications}.\hskip 1em
  plus 0.5em minus 0.4em\relax Springer, 2016, pp. 179--187.

\bibitem{zhao2016loss}
H.~Zhao, O.~Gallo, I.~Frosio, and J.~Kautz, ``Loss functions for image
  restoration with neural networks,'' \emph{IEEE Transactions on computational
  imaging}, vol.~3, no.~1, pp. 47--57, 2016.

\bibitem{zbontar2018fastmri}
J.~Zbontar, F.~Knoll, A.~Sriram, M.~J. Muckley, M.~Bruno, A.~Defazio,
  M.~Parente, K.~J. Geras, J.~Katsnelson, H.~Chandarana \emph{et~al.},
  ``fast{MRI}: An open dataset and benchmarks for accelerated {MRI},''
  \emph{arXiv preprint arXiv:1811.08839}, 2018.

\bibitem{Tieleman2012}
T.~Tieleman and G.~Hinton, ``{Lecture 6.5---RMSProp: Divide the gradient by a
  running average of its recent magnitude},'' COURSERA: Neural Networks for
  Machine Learning, 2012.

\bibitem{falcon2019pytorch}
W.~Falcon, ``Pytorch lightning,'' \emph{GitHub. Note:
  https://github.com/PyTorchLightning/pytorch-lightning Cited by}, vol.~3,
  2019.

\bibitem{ulyanov2018deep}
D.~Ulyanov, A.~Vedaldi, and V.~Lempitsky, ``Deep image prior,'' in
  \emph{Proceedings of the IEEE Conference on Computer Vision and Pattern
  Recognition}, 2018, pp. 9446--9454.

\bibitem{van2018compressed}
D.~Van~Veen, A.~Jalal, M.~Soltanolkotabi, E.~Price, S.~Vishwanath, and A.~G.
  Dimakis, ``Compressed sensing with deep image prior and learned
  regularization,'' \emph{arXiv preprint arXiv:1806.06438}, 2018.

\bibitem{liu2019image}
J.~Liu, Y.~Sun, X.~Xu, and U.~S. Kamilov, ``Image restoration using total
  variation regularized deep image prior,'' in \emph{ICASSP 2019-2019 IEEE
  International Conference on Acoustics, Speech and Signal Processing
  (ICASSP)}.\hskip 1em plus 0.5em minus 0.4em\relax IEEE, 2019, pp. 7715--7719.

\bibitem{abadi2016tensorflow}
M.~Abadi, P.~Barham, J.~Chen, Z.~Chen, A.~Davis, J.~Dean, M.~Devin,
  S.~Ghemawat, G.~Irving, M.~Isard \emph{et~al.}, ``Tensorflow: a system for
  large-scale machine learning.'' in \emph{OSDI}, vol.~16, 2016, pp. 265--283.

\bibitem{kingma2014adam}
D.~P. Kingma and J.~Ba, ``Adam: {A} method for stochastic optimization,''
  \emph{arXiv preprint arXiv:1412.6980}, 2014.

\end{thebibliography}
\bibliographystyle{IEEEtran}

\end{document}


\title{On hallucinations in tomographic \\ image reconstruction \\ (Supplementary Material)}
\author{Sayantan Bhadra, Varun A. Kelkar, Frank J. Brooks and Mark A. Anastasio,~\IEEEmembership{Senior Member,~IEEE}}
\maketitle

\section{Sampling mask}
\begin{figure}[!htb]
\centering
\centerline{\includegraphics[width=0.3\linewidth]{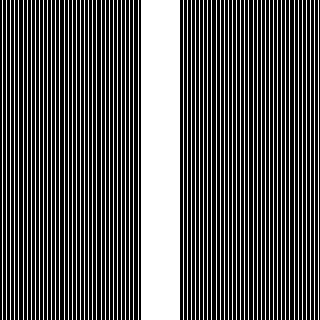}}
\caption{Sampling mask}
\label{fig:mask}
\end{figure}

A uniform Cartesian undersampling mask was employed in our simulation studies for generating the incomplete k-space data described in Sec. IV-A of the manuscript. The full k-space data were undersampled by a factor of 3 using the uniform sampling mask shown in Fig. S. \ref{fig:mask}.

\vspace{-0.5cm}

\section{Details of reconstruction methods}
\label{sec:recon}
The details of the three reconstruction methods employed in our studies are presented below. They are based on the imaging model:
\begin{equation}\label{eq:imaging_DD}
\vec{g} = \vec{H}\coeff + \vec{n},
\end{equation}
where $\coeff \in \mathbb{E}^N$ is the sought-after coefficient vector, $\vec{g} \in \mathbb{E}^M$ is the observed measurement data, $\vec{H} \in \mathbb{E}^{M \times N}$ is the system matrix, and $\vec{n} \in \mathbb{E}^N$ is iid Gaussian noise. 

\subsection{Penalized least squares with TV regularization (PLS-TV)}
\label{sec:plstv}
The PLS-TV method involves solving the penalized least-squares optimization framework with the penalty term as the TV penalty:
\begin{equation}\label{eq:optim}
\hat{\coeff} = \argmin_{\coeff} ||\vec{g}-\vec{H}\coeff||^2_2 + \lambda ||\coeff||_{TV},
\end{equation}
where $\lambda$ is the regularization parameter. Proximal gradient methods are commonly employed to implement the PLS-TV method \cite{chambolle2004algorithm, beck2009fast, afonso2010augmented}. In this study, PLS-TV reconstruction was performed for a dataset of measurements corresponding to 69 different images using the Berkeley Advanced Reconstruction Toolbox (BART) \cite{uecker2015berkeley,BART}. BART performs PLS-TV reconstruction using the augmented Lagrangian based optimization method proposed in \cite{afonso2010augmented}. The regularization parameter $\lambda$ in Eq. \eqref{eq:optim} was chosen by first performing image reconstruction on a subset of the dataset, with different values of $\lambda$. The value of $\lambda$ which provided the lowest mean of the root mean squared error (RMSE) metric over the subset was chosen, and used for image reconstruction of all the images in the dataset.

\subsection{U-Net-based reconstruction}
The following procedure, known as \textit{image-domain learning}, was employed for the U-Net based image reconstruction. First, initial estimates of the images were obtained from the measurement data by use of the pseudoinverse operator. They were then employed as inputs to a convolutional neural network (CNN), which was trained in order to produce artifact-free images, similar to images from the ground truth distribution \cite{jin2017deep, han2018framing, hyun2018deep, antholzer2019deep}. As is common practice, the CNN architecture used in this study is the U-Net \cite{ronneberger2015unet}. A U-Net consists of two CNNs that represent a downsampling path followed by an upsampling, path respectively, and skip connections \cite{drozdzal2016importance} between similar levels in the downsampling and upsampling paths. Let the initial estimate from the measurement data be denoted as $\coeff'$ and the function computed by the U-Net be represented as $B(\coeff'; w)$ where $B: \mathbb{E}^N \rightarrow \mathbb{E}^N$ and $w$ denotes the weight parameters of the U-Net. Given a training data set of initial estimate-ground truth pairs  $\big\{\coeff'_i,\coeff_i \big\}_{i=1}^{D}$ where $D$ is the size of the training data set, the optimal weight parameters $w^*$ are learned by approximately solving the following optimization problem:
\begin{equation}\label{eq:training}
w^* = \argmin_w \sum_{i=1}^{D} \mathcal{L}\Big(B(\coeff'_i,w),\coeff_i\Big),
\end{equation}
where $\mathcal{L}(\cdot,\cdot)$ is a suitable loss function. In this work, mean absolute error was used as the loss function \cite{zhao2016loss}. 
The model for the U-Net was based on the single-coil baseline U-Net architecture provided in \cite{zbontar2018fastmri}. A stochastic gradient-based method known as RMSProp \cite{Tieleman2012} was employed to solve the optimization problem in Eq. \eqref{eq:training}. After this iterative scheme for training the U-Net reached convergence, the trained U-Net was used to reconstruct images from a previously unseen test measurement dataset, where an initial estimate $\coeff'_{test}$ computed from a test measurement data was employed to obtain the reconstructed image $\hat{\coeff}_{test} = B(\coeff'_{test},w)$.
The training and testing of the U-Net based reconstruction was performed using code available at \url{https://github.com/facebookresearch/fastMRI}, which utilizes PyTorch Lightning \cite{falcon2019pytorch}.

\subsection{Deep image prior (DIP)}
Recently, Ulyanov \textit{et al.}\cite{ulyanov2018deep} showed that a CNN $G : \mathbb{R}^k \rightarrow \mathbb{E}^N$ with randomly initialized weights $w$ and random input $\vec{z} \in \mathbb{R}^k$ can be an effective regularizer for image restoration problems such as denoising, super-resolution and inpainting. This method of regularization, known as deep image prior (DIP), utilizes the observation that the structure of deep convolutional networks captures several low-level image statistics and is biased towards smooth, natural images. Van Veen \textit{et al.} \cite{van2018compressed} extended the DIP framework to applications in tomographic imaging from incomplete measurements with encouraging results. Essentially,
image reconstruction using the DIP method can be formulated in terms of the following optimization problem:
\begin{align}\label{eq:DIP}
w^* &= \argmin_{w} ||\vec{g}-\vec{H}G(\vec{z};w)||^2_2, \nonumber\\
\hat{\coeff} &= G(\vec{z}; w^*)
\end{align}
where $\vec{z}$ and $w$ are randomly initialized.

It has been shown in \cite{ulyanov2018deep,van2018compressed} that the DIP method overfits the measurement noise upon convergence. Hence, further regularization may be required, either in the form of early stopping or with the addition of penalties in the optimization problem in Eq. \eqref{eq:DIP}. Inspired by \cite{liu2019image}, in our experiments, image reconstruction using the DIP method with TV regularization (DIP-TV) was performed by approximately solving the following optimization problem:
\begin{align}\label{eq:DIP-TV}
w^* &= \argmin_w ||\vec{g}-\vec{H}G(\vec{z},w)||^2_2 + \lambda ||G(\vec{z},w)||_{\rm TV},\nonumber\\ 
\hat{\coeff} &= G(\vec{z}; w^*)
\end{align}
where $\vec{z}$ and $w$ were randomly initialized, and $\lambda$ is the regularization parameter. The same U-Net architecture employed for the U-Net based reconstruction was employed for DIP-TV, and was implemented in TensorFlow \cite{abadi2016tensorflow}. Similar to the implementation of the PLS-TV method as outlined in Sec. \ref{sec:plstv}, the regularization parameter $\lambda$ for the TV penalty in Eq. \eqref{eq:DIP-TV} was chosen by first performing image reconstruction on a subset of the dataset, with different values of $\lambda$. 
Subsequently, the value which provided the lowest mean RMSE over the subset was chosen to perform image reconstruction from all the measurements. The optimization problem in Eq. \eqref{eq:DIP-TV} was approximately solved using a stochastic gradient algorithm called Adam \cite{kingma2014adam}.

\section{Examples of measurement space hallucination maps}
While the results in our simulation studies focused on the effect of null space hallucination maps, measurement space hallucination maps can also be computed corresponding to reconstructed images from different methods. The measurement space hallucination map is denoted as
\begin{equation}\label{eq:hal_map_meas}
    \hcmeas^{HM} \equiv \hcmeas-\hcpinv,
\end{equation}
which describes the consistency between the measurement component of the reconstructed image $\hcmeas$ with respect to the truncated pseudoinverse solution $\hcoeff_{tp}$ that can be stably obtained from the measurement data $\vec{g}$. Furthermore, unlike the null space hallucination map, the computation of $\hcmeas^{HM}$ does not require the knowledge of the true object $\vec{\theta}$. On the other hand, the error map between $\hcmeas$ and $\coeff_{meas}$ is a similar but different error quantity that lies in the measurement space $\Umeas$ and requires the knowledge of $\coeff$:
\begin{equation}\label{eq:e_meas}
    \hcmeas^{EM} \equiv \hcmeas - \cmeas.
\end{equation}
In some cases, $\hcmeas^{HM}$ and $\hcmeas^{EM}$ may not convey the same information due to the differences that can exist between $\cmeas$ and $\hcpinv$. These differences arise when there is significant measurement noise in the imaging system or due to modeling error in $\vec{H}$, or both. With the \enquote{true} imaging operator denoted as $\tilde{\vec{H}}$ and the assumed imaging operator as $\vec{H}$, Eq. \eqref{eq:imaging_DD} can be re-written as 
\begin{equation}
    \vec{g} = \tilde{\vec{H}}\coeff + \vec{n}
\end{equation}
Accordingly, $\hcpinv$ can be expressed as
\begin{equation}\label{eq:pinv_noise_true}
    \hcpinv = \MP_P \vec{g} \approx \MP_P (\tilde{\vec{H}}\coeff+\vec{n}) = \MP_P\tilde{\vec{H}}\coeff + \MP_P \vec{n}. 
\end{equation}
On the other hand, $\cmeas$ is represented as
\begin{equation}\label{eq:theta_meas}
    \cmeas \equiv \MP_P \vec{H}\coeff.
\end{equation}
It can be observed from Eq. \eqref{eq:pinv_noise_true} and Eq. \eqref{eq:theta_meas} that, when either or both of the quantities $||\vec{n}||_2^2$ and $||\vec{H}-\tilde{\vec{H}}||_2^2$ is non-trivial, $||\hcpinv-\cmeas||_2^2 $ is likely to be significant. In such cases, $\hcmeas^{HM}$ and $\cmeas$ may represent different information. 
 
\begin{figure*}[!t]
    \centering
     \includegraphics[width=\linewidth]{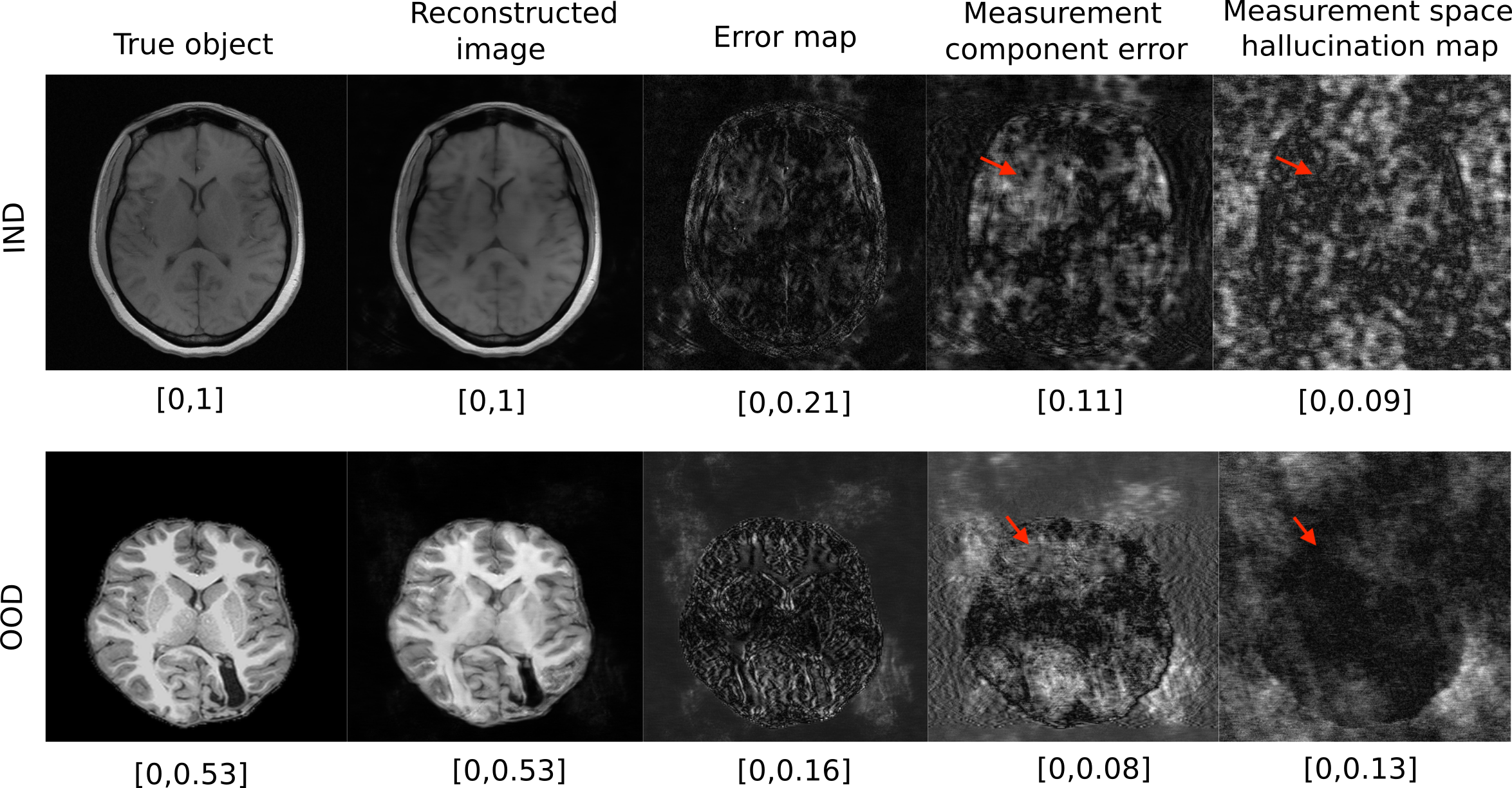}
      \caption{Measurement space hallucination maps for reconstructed images using the U-Net method corresponding to an IND (above) and an OOD (below) object. Note that the measurement component error map and the measurement space hallucination map are appreciably different. The red arrows point towards a region in each type of object where such differences can be clearly seen.
      }
      \label{fig:meas_hm}
\end{figure*}

Figure S.\ref{fig:meas_hm} shows examples of measurement space hallucination maps for images reconstructed by the U-Net method corresponding to an in-distribution (IND) object and an out-of-distribution (OOD) object. It can be observed that the measurement component error map has appreciable differences compared to the measurement space hallucination map. These differences can be attributed to the presence of non-trivial measurement noise as well as additional phase noise disturbance during simulation of the k-space data resulting in model error, since the phase noise is unknown and assumed to be zero during reconstruction. The measurement space hallucination maps as shown here may provide an insight into how well a given reconstruction method maintains data consistency and be compared with other reconstruction methods.

\bibliography{DLRHM}{}
\bibliographystyle{IEEEtran}